\documentclass[%
 reprint,
superscriptaddress,
showpacs,preprintnumbers,showkeys,
 amsmath,amssymb,
 aps,
prb,
floatfix,
]{revtex4-1}

\usepackage{amsmath}
\usepackage{array,longtable}
\usepackage{natbib}
\usepackage{graphicx}
\usepackage{color}
\usepackage{transparent}
\graphicspath{{Figs/}}
\usepackage{longtable}
\usepackage{latexsym}
\usepackage{booktabs}
\usepackage{ctable}
\usepackage{braket}
\usepackage[caption=false]{subfig}
\usepackage{url}
\usepackage{hyperref}
\newcommand*{\gga}[1]{#1}

\newcommand*{\e}[1]{{\mathrm e}^{#1}}

\bibpunct{}{}{,}{s}{}{}
\begin{document}
\author{Mohsen Sotoudeh}
\email{mohsen.sotoudeh@uni-ulm.de}
\affiliation{Institute of Theoretical Chemistry, Ulm University, 
Albert-Einstein-Allee 11, 89081 Ulm, Germany}
\author{Marian David Bongers-Loth}
\affiliation{Institut f\"ur Materialphysik, 
Universit\"at G\"ottingen, Friedrich-Hund-Platz 1,
  37077 G\"ottingen, Germany}
\author{Vladimir Roddatis}
\affiliation{Institut f\"ur Materialphysik, 
Universit\"at G\"ottingen, Friedrich-Hund-Platz 1,
  37077 G\"ottingen, Germany}
\affiliation{Helmholtz Centre Potsdam,
GFZ German Research Centre for Geosciences
Telegrafenberg, 14473 Potsdam, Germany}

  \author{Jakub \v{C}\'{i}\v{z}ek}
\affiliation{Department of Low-temperature Physics, Charles University
  in Prague, V Hole\v{s}ovi\v{c}k\'{a}ch 2, 180 00 Praha 8, Czech
  Republic}
\author{Carsten Nowak}
\affiliation{Institut f\"ur Materialphysik, 
Universit\"at G\"ottingen, Friedrich-Hund-Platz 1,
  37077 G\"ottingen, Germany}
\affiliation{XLAB, Universit\"at G\"ottingen, Justus-von-Liebig-Weg 8, 37077 G\"ottingen, Germany}
\author{Martin Wenderoth}
\affiliation{IV. Physikalischen Institut, Universit\"at G\"ottingen, 
  Friedrich-Hund-Platz 1,
  37077 G\"ottingen, Germany}
\author{Peter Bl\"ochl}
\email{peter.bloechl@tu-clausthal.de}
\affiliation{Institute for Theoretical Physics, Clausthal University
 of Technology, Leibnizstr. 10, 38678 Clausthal-Zellerfeld, Germany}
\affiliation{Institut f\"ur Theoretische Physik, 
 Universit\"at G\"ottingen, Friedrich-Hund-Platz 1,
 37077 G\"ottingen, Germany}
\author{Astrid Pundt}
\email{astrid.pundt@kit.edu}
\affiliation{Institut f\"ur Materialphysik, 
Universit\"at G\"ottingen, Friedrich-Hund-Platz 1,
  37077 G\"ottingen, Germany}
\affiliation{Institut f\"ur Angewandte Materialien-Werkstoffkunde,
Karlsruher Institut f\"ur Technologie, Kaiserstr. 12, 76131 Karlsruhe, 
Germany}

\title{Hydrogen related defects in titanium dioxide at the 
  interface to palladium}
\date {\today}

\begin{abstract}
A metal oxide support and a catalytically active metal are the two main 
ingredients for complex catalysts used in heterogeneous catalysis. The gas 
environment can change the catalyst during the reaction, modifying its 
structural and electronic properties. Here, we use monochromated electron energy 
loss spectroscopy (EELS)  to reveal hydrogen-pressure-dependent changes of the 
electronic structure at the Pd/rutile-TiO$_2$ interface in an environmental 
transmission electron microscope (ETEM). Hydrogen-induced changes are observed 
in rutile-TiO$_2$ within $2$~nm from  the interface at $10$~Pa of hydrogen 
pressure, in the Ti $L_{3,2}$ EEL spectra. Lower pressures such as $1$~Pa show 
no changes in the EEL spectra. We attribute the observed changes in the EEL 
spectra to hydrogen-induced defects accumulating in the vicinity of the 
interface. Based on DFT calculations, we developed a thermodynamic  multistate 
defect (TMD) model of the interface and the bulk of the rutile-TiO$_2$. This TMD 
model predicts high concentrations of positively charged defects accumulating at 
the interface. The presence of the Schottky barrier stabilizes these defects by 
significantly lowering their formation energy. Our findings provide important 
new insights into catalytic processes taking place at metal/metal oxide 
interfaces in hydrogen gas environments.
\end{abstract}


\maketitle

\section{Introduction}
Metal/oxide systems using titanium dioxide (TiO$_2$) as a support have 
been used in a broad range of technological applications where they 
serve as model systems for fundamental process studies in surface 
science~\cite{diebold2003surface, diebold2010oxide},
catalysis~\cite{tauster1981strong, roland1997nature,
park2015role, Nowotny_2008, ni2007review, waser2009redox, ma14_cr_114_9987,
bowker2002co, sa2007imaging, liu2011advanced, chen2015black, Ola201516, 
Burlaka_2017, Roddatis_2018}, the field of resistive 
switching~\cite{mohammad2016state, moballegh2015electric, kwon2010atomic} 
as well as for Schottky contacts~\cite{kobayashi1994reactions,
cerchez2013dynamics, strungaru2015interdependence, roland1995investigations}. 
One central question in these studies is how chemical and physical properties 
of the metal/TiO$_2$ interface are influenced by an electric 
field~\cite{kwon2010atomic}, by metal-support interaction~\cite{sa2007imaging} 
or by defect concentrations~\cite{cerchez2013dynamics, 
strungaru2015interdependence}. The latter can be also influenced by the environment, 
e.g. by the presence of hydrogen, as utilized in for catalysts and gas sensors.

Defects, such as oxygen vacancies, 
Ti vacancies, and Ti interstitials, and their properties in bulk 
rutile-TiO$_2$ (r-TiO$_2$) have been experimentally 
studied~\cite{diebold2003surface, Nowotny_2008, grant1959properties,
nowotny2006electricalI, nowotny2006electricalII, nowotny2006electricalIII,
nowotny2006electricalIV, nowotny2011effect, bak2003defectI, bak2003defectII, 
bak2003defectIII}. On the other side, the hydrogen concentration in pristine 
r-TiO$_2$ crystals can be found~\cite{Johnson_1973} between 
10$^{17}$~cm$^{-3}$ and 10$^{19}$~cm$^{-3}$. Interstitial
hydrogen~\cite{Peacock_2003, Filippone_2009} and
hydrogenated oxygen vacancy~\cite{Filippone_2009} (H$_{\text{O}}$)
are detected by infrared spectroscopy~\cite{Herklotz_2011} 
and discussed as two important hydrogen-related defects, 
which act as dopants in r-TiO$_2$.

According to recent theoretical studies on bulk TiO$_2$~\cite{Filippone_2009, 
Bjorheim_2013,aschauer12_pccp14_16595, sotoudeh14_aip4_027129,
sotoudeh14_epjap67_30401, mo15_sr5_17634}, 
the favorable adsorption sites for H atoms are
oxygen vacancies (V$_{\text{O}}$), and, in the absence of vacancies, 
interstitial sites (H$_{\text{i}}$). All these point defect studies were 
carried out using first-principles calculations based on density-functional 
theory (DFT).

Palladium (Pd) or platinum (Pt) are often used as metal partners. 
They are highly active in
hydrogenation reactions~\cite{wells1998platinum}, and
for dissociative chemisorption of H$_2$ molecules at their surfaces
\cite{Wicke_1978,christmann98_surfsci395_182,rieder99_jchemphys110_559}.
Pd is a well-studied catalyst and also a system for hydrogen
ad- and absorption~\cite{teschner2008roles, Brandt2009191, 
sandrock1992hydrogen}. Hydrogen solves on interstitial sites in the 
Pd lattice  in the Pd-H $\alpha$-phase and the Pd-H hydride phase 
\cite{Wicke_1978,manchester1994h}. The solubility of H depends
on the hydrogen chemical potential~\cite{Sievert1929}. It is, for thin 
films, also affected by the film's  
microstructure~\cite{flanagan1991palladium, pundt2006hydrogen, 
kirchheim2007reducingI, kirchheim2007reducingII} and the mechanical stress
state~\cite{wagner2016quasi}. For low chemical potentials, the 
hydrogen-related mechanical stress is small~\cite{wagner2016mechanical}. 
For high chemical potentials of H, the mechanical stress can be in the 
GPa-range~\cite{pundt2006hydrogen}. The room-temperature kinetics of 
H in Pd is fast~\cite{mehrer2007diffusion}. The H diffusion coefficient of
$3.2\times10^{-11}$~m$^2$/s leads to diffusion over $\mu$m distances
within seconds. For a Pd thin film, this allows for fast equilibration and, 
further, to fast transport of atomic hydrogen to the film/TiO$_2$ interface.

At the Pd/TiO$_2$ interface, the energetic difference 
between the workfunction of Pd and the electron affinity 
of TiO$_2$ results in electron transfer. A
space charge layer builds up and a Schottky-barrier 
establishes at the contact~\cite{moench2004electronic, tung2014physics} 
according to Schottky's classical picture. Hydrogen 
diffuses through the Pd to the interface where it spills over into 
the TiO$_2$ and affects the Schottky 
barrier thickness and/or height~\cite{cerchez2013dynamics}. 
Irreversible changes in the conductivity of hydrogen loaded Pd/TiO$_2$
hint on the presence of stable hydrogen-oxygen vacancy 
complexes~\cite{strungaru2015interdependence}.
Hydrogen additionally lowers the critical field necessary 
for electroforming processes and can eventually lead to the formation of 
substoichiometric phases in the titanium 
dioxide~\cite{strungaru2015interdependence}.
Recent experiments reveal additional local (spatial) changes of
the Schottky barrier height that could be attributed to the formation
of substoichiometric Magn{\'e}li phases (Ti$_n$O$_{2n-1}$, $n=4$ to $10$) 
in the vicinity of the interface~\cite{strungaru2015interdependence}.
All suggested interpretations~\cite{paulose2005unprecedented, 
 strungaru2015interdependence} 
of the observed Schottky barrier height changes hint on a dominant 
contribution of defects in the near-interface region in the TiO$_2$.
The microscopic details of the interface changes are still under 
debate. However, a strong contribution of defects is beyond question.
The dominant type of hydrogen-related defects in r-TiO$_2$ 
is still unknown, especially in the close vicinity to the Pd interface.

In this study, we determine hydrogen-related changes in r-TiO$_2$ in the 
nanometer vicinity of the Pd/r-TiO$_2$ interface. Experimentally, we use 
electron energy-loss spectroscopy (EELS) in an environmental TEM (ETEM). 
First-principle calculations based on the generalized gradient approximation 
(GGA) of the DFT were carried out to assess the influences of different defects 
on the electronic density of states (DoS). Results originating from these 
calculations serve as input parameters for a thermodynamic 
multistate defect model (TMD model). This TMD model allows to implement the 
experimental conditions and to gain information about defect concentrations in 
the r-TiO$_2$ and at the interface to the metal.

\section{Methods}
\label{chap:methods}
\subsection{Pd/TiO{$_2$} sample preparation and characterization}
Pd($111$) films were deposited on pretreated r-TiO$_2$($110$)
crystals (CrysTec GmbH, Berlin, epi-polished), by magnetron sputter 
deposition in a high vacuum system. The r-TiO$_2$ substrates 
were etched in $20$~\% fluoric acid, rinsed in purified water and 
annealed for $1$~h at $1173$~K in an O$_2$ gas
stream of approximately $10^{5}$~Pa. Such pretreatment results in 
the formation of clean and smooth well-defined surface terraces on 
r-TiO$_2$ as verified by x-ray photoelectron spectroscopy (XPS) and 
atomic force microscopy (AFM). Subsequently, the Pd films were 
deposited at $1024$~K in an oxygen atmosphere with a pressure 
of $7 \times 10^{-3}$~Pa and a sputter rate of approximately 
$11.8$~nm/min. After film deposition, the samples were cooled 
down in the oxygen atmosphere to about $373$~K. 
The chosen deposition conditions resulted in Pd grain sizes above
$1$~$\mu$m. Further details are provided elsewhere~\cite{Bongers_2018}.

Cross-sectional specimens required for high-resolution transmission 
electron microscopy (HRTEM) and EELS studies are prepared by 
mechanical polishing and subsequent low-angle (6$^{\circ}$) Ar$^+$ 
ion milling at $3$~kV in PIPS $691$ (Gatan). 
At the final stage of ion milling, the acceleration voltage was 
gradually decreased to $0.5$~kV according to the procedure described by
S\"{u}ess \textit{et al.}~\cite{suess2011minimization}. 

Variable energy positron annihilation spectroscopy (VEPAS) studies on
Pd/r-TiO$_2$ reveal no detectable change of
the Ti-vacancy concentration in r-TiO$_2$ in the here addressed
pressure range. VEPAS probes for
vacancy concentrations between $10^{-7}$ to $10^{-4}$ per
atom~\cite{krause1999positron}. Further, the concentration of the related
Ti-interstitials is, here, regarded as low, with respect to the
concentration of oxygen vacancies, under the chosen preparation
conditions~\cite{bak2003defectII} and assumed to be constant in the
experimentally studied pressure range. Therefore, we do not consider
any Ti-related defects in our simulations. Details on the VEPAS
measurements are presented in Appendix~\ref{chap:appex_VEPAS}.

\subsection{Electron energy-loss spectroscopy (EELS) experimental
details}
The experiments on the Pd/TiO$_2$ interfaces were performed using an 
FEI Titan $80$-$300$ ETEM operated at $300$~keV. The ETEM is equipped
with a Cs-image corrector, monochromator and GATAN image filter 
(GIF) Quantum $965$. TEM experiments were performed in a high vacuum 
($10^{-5}$~Pa) and at H$_2$ gas pressures up to $10$~Pa. The sample was
equilibrated in H$_2$ for 1~hour at each pressure. 
During this time the electron beam in the ETEM was blanked.
The hydrogen loaded Pd can be 
regarded as an atomic hydrogen source for this study. Spectrum images (SI) 
were collected at different places in the close nanometer-range vicinity of the 
Pd/TiO$_{2}$ interface, at selected H$_2$ gas pressure ranging from $10^{-1}$~Pa
to $10$~Pa. To reveal changes in the Ti $L$ electron energy-loss near-edge 
structure (ELNES) of about $100$~meV an advanced procedure was used
that is described in details in Appendix~\ref{chap:appex_EELS_procedure}.

\subsection{Computational details}
\label{sec:computationaldetails}
First-principles calculations based on density-functional theory
(DFT)~\cite{hohenberg64_pr136_B864,kohn65_pr140_1133} were performed 
on pristine r-TiO$_2$ and r-TiO$_2$ containing selected defects.
We used the generalized gradient approximation (GGA) functional
PBE~\cite{perdew96_prl77_3865}. 

The calculations were carried out with the Projector Augmented Wave
(PAW)~\cite{bloechl94_prb50_17953} method using the CP-PAW code~\cite{cp-paw}. 
The augmentation has been set up using the systematic projector
construction~\cite{bloechl12_arxiv1210_5937}. For the augmentation, a
$s^{2}p^{2}d^{1}$ ($3s$, $3p$, $3d$, $4s$, and $4p$ orbitals) set of projector 
functions was used for the Ti atom, and a $s^{2}p^{1}d^{1}$ ($2s$, $2p$, $3s$, 
and $3d$ orbitals) set for the O atom. The superscripts denote the
number of projector functions per angular momentum channel. For the Ti
atom, the $3s$ and $3p$ core electrons were treated on the same level
as the valence electrons.

A plane wave cutoff of 40~Ry has been chosen for the wave functions
and 80~Ry for the charge density. The Brillouin-zone integration has
been performed with the linear tetrahedron
method~\cite{jepsen71_ssc9_1763,lehmann72_pssb54_469} and the
so-called Bl\"{o}chl corrections~\cite{bloechl94_prb49_16223}. A
5$\times$5$\times$8 k-point mesh has been used for the unit cell
containing two formula units. Defect calculations were simulated by
2$\times$2$\times$3 supercells with 72 atoms plus the defect.  All
atomic positions are optimized without symmetry constraints. 
A theoretical lattice constant 4.562~{\AA} of bulk r-TiO$_2$ was obtained, 
in good agreement with the experimental~\cite{Cromer_1955} value of 4.593~{\AA}. 
The calculated $c/a$ ratio matches the experimental~\cite{Cromer_1955} ratio of $0.644$.
This lattice constant was used for the supercells. Collinear
spin polarization was allowed in all calculations.

Three different defect types in r-TiO$_2$ were selected: the oxygen vacancy
V$_{\text{O}}$, the interstitial hydrogen H$_{\text{i}}$ and the
oxygen vacancy filled with hydrogen H$_{\text{O}}$. The interstitial
titanium Ti$_{\text{i}}$ was not considered as the concentration was
found experimentally not to depend on the hydrogen partial pressure
(Appendix~\ref{chap:appex_VEPAS}).

Crystal field splitting divides the Ti-$d$ orbitals, located at 
the center of a regular TiO$_6$ octahedron, into high-lying, doubly degenerate 
$e_g$ and low-lying triply degenerate $t_{2g}$ orbitals. Note that the 
crystal-field splitting originates from the covalent interaction with the oxygen 
neighbors. The stronger $\sigma$ bond of the $e_g$ states in comparison to the 
$\pi$ bonds formed by the $t_{2g}$ states, shifts the antibonding $e_g$ states 
energetically above the $t_{2g}$ states.
 
\begin{figure}[!ht]
\centering
  \includegraphics[clip,width=0.8\columnwidth]{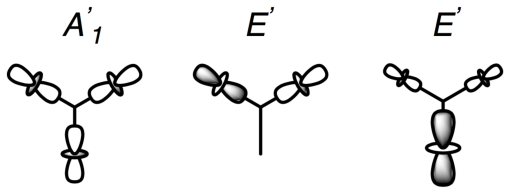}%
  \caption{\label{fig:orbital-rep}Corresponding Ti-$d$ orbitals 
  representing the local symmetry at
  an O-site (defect center), in the ($110$) plane. The Ti-$d$
  states pointing towards the defect center are dominated by the
  approximate D$_{3\mathrm{h}}$ symmetry of the defect. They are
  classified as $A'_1$ and $E'$. $E'$ appears in two degenerate modifications. 
  For the other $d$ orbitals on the
  Ti-neighbors of the defect, we refer to the approximate octahedral
  symmetry (O$_{\text{h}}$).}
\end{figure}

The molecular orbitals associated with defect electronic states are related to 
the surrounding Ti $d$-orbitals. Linear combinations of these orbitals result in 
bonding and antibonding states. The $e_g$ orbitals on the Ti neighbors of the 
defect can be divided into two situations: the situation where the orbitals 
point towards the defect center. They can be superimposed into one $A'_1$ 
orbital and two $E'$ orbitals, as shown in Fig.~\ref{fig:orbital-rep}. In the 
second situation, the orbitals point orthogonal to this direction. It is denoted 
as  $\delta e_g$. This situation appears not to be affected by the presence of 
defects, as will be shown later.

Furthermore, we considered the Ti-$t_{2g}$, which are classified
according to the approximate octahedral symmetry of the Ti-site.

In order to represent the Ti-$d$ states pointing towards the defect center, 
the local $z$-axis for the Ti-$d$ orbitals has been
defined by the Ti-O bond pointing away from the defect center. A
second, approximately orthogonal TiO-bond has been used to define the
$xz$-plane.

The formation energy of defect $X_\sigma$ is given by the defect energy 
$E_\sigma$ and the chemical potentials for electrons, oxygen and hydrogen atoms.
\begin{eqnarray}
E_{f}[X_\sigma]=E_\sigma-\eta_{O,\sigma}\mu_O-\eta_{H,\sigma}\mu_H-\eta_{e,\sigma}\mu_e
\end{eqnarray}
The defect energies $E_\sigma$ and stoichiometric factors $\eta_{O,\sigma}$, 
$\eta_{O,\sigma}$, and $\eta_{H,\sigma}$ are given below in 
table~\ref{tab:pottstate}. The defect energy $E_\sigma$ is calculated as the 
energy difference $E[\mathrm{TiO}_2:X_\sigma]-E[\mathrm{TiO}_2]$ of the 
supercell with and without the defect. The lattice constants of the supercell 
have been taken from the calculations of bulk r-TiO$_2$, provided in 
table~\ref{tab:tio2-strc}.

In order to avoid large values without physical significance, we introduce  
reference values for each of the three chemical potentials. We choose 
the valence band top $\mu_{e}^{ref}=\epsilon_v$ for the electron 
chemical potential and one-half of the calculated molecular energies 
$\mu_{O}^{ref}=\frac{1}{2}E[\mathrm{O}_2]$ and 
$\mu_{H}^{ref}=\frac{1}{2}E[\mathrm{H}_2]$ for oxygen and hydrogen atoms, 
respectively.
 
In our supercell, the defects in r-TiO$_2$ produce a deviation of the 
valence-band maximum and conduction-band minimum ranging between ${+}0.15$~eV 
and ${-}0.02$~eV from the calculated bulk values. In our calculations, we 
measure the electron energies relative to the valence-band top of the perfect 
r-TiO$_2$ crystal. 

The thermodynamic charge state level
\begin{eqnarray}
\epsilon(q_1/q_2)=\frac{E[X^{q_2}]-E[X^{q_1}]}{q_2-q_1}
\label{eq:thermochargestatelevel}
\end{eqnarray}
defines the position of the electron chemical potential $\mu_e$ for which 
two charge states ($q_1$ and $q_2$) coexist in the same quantity. For lower 
$\mu_e$ the more positive charge state is thermodynamically stable and 
for higher $\mu_e$ the more negative charge state is thermodynamically stable. 
$\mu_e$ is commonly
referred to as the Fermi level.

\section{Results}
\subsection{EELS experiments}
\label{sec:eels-exp}
\begin{figure}[!ht]
\centering
\subfloat[]{%
 \includegraphics[clip,width=0.55\columnwidth]{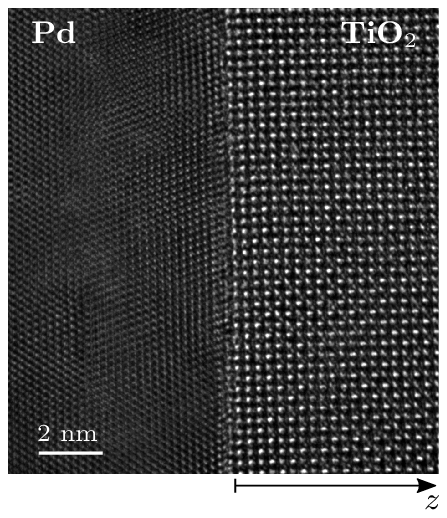}
}
 
\subfloat[]{%
 \includegraphics[clip,width=1.0\columnwidth]{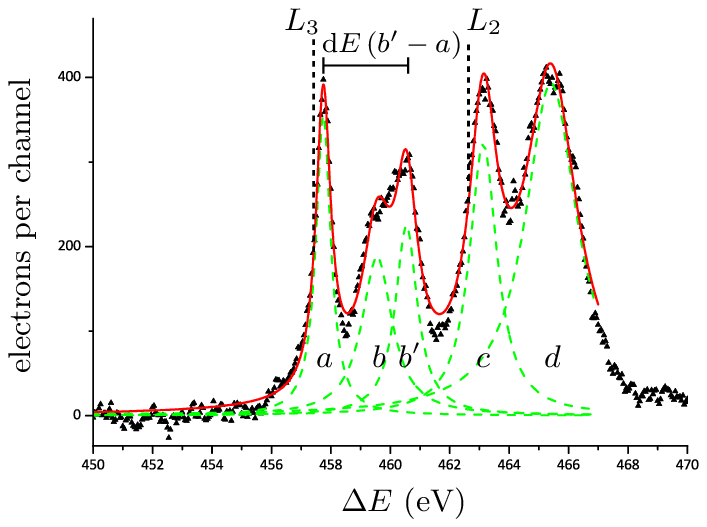}
 \label{fig:Ti-L-ELNES_with_fit}
 }
\caption{\label{fig:Pd-TiO2_interface}(a) HRTEM picture of an as 
prepared Pd/r-TiO$_2$ interface. A 
sharp interface between one to two monolayers thickness is revealed and no one 
or higher dimensional defects can be found in the interface 
region. (b) Typical  Ti $L_{3\textrm{,}2}$ 
electron loss near edge structure (ELNES) (triangles)
after a Hartree-Slater background (HSB) correction.
Fitting with five Lorentz functions (green dashed lines) yields
an envelope function (red line) which describes
the features in the Ti $L_{3\textrm{,}2}$ ELNES appropriately well. 
Indicated are the four white lines ($a$, $b$/$b'$, $c$, $d$) as well
as the energy splitting d$E(b'-a)$. $\Delta E$ is the 
measured energy-loss. This spectrum was detected in the as prepared state, 
for all distances.}
\end{figure}

Fig.~\ref{fig:Pd-TiO2_interface} shows a HRTEM-image confirming a flat and sharp 
interface between the Pd film and the r-TiO$_2$ substrate.
EEL spectra were collected in r-TiO$_2$ in the $0-20$~nm vicinity to the 
interface at different H$_2$ gas pressure below $1$~Pa. Details about the 
spectra evaluation are given in Appendix~\ref{chap:appex_EELS_procedure}, 
Fig.~\ref{fig:EELS_experiment}. For 
different H$_2$ pressures below 1 Pa, all EEL spectra show similar results, 
reflecting the bulk spectrum of r-TiO$_2$ shown in 
Fig.~\ref{fig:Ti-L-ELNES_with_fit}. Fig.~\ref{fig:Ti-L-ELNES_with_fit} focuses 
on the Ti $L$ edge. Any valence change or local symmetry change due to the 
presence of defects leads to modifications of the system's energy levels, which 
is especially measurable in EELS on the Ti $L$ 
edge~\cite{moballegh2015electric, stoyanov2007effect}.
The Ti $L_{3,2}$ edge originates from excitation of Ti-2$p$ state electrons 
into the unoccupied states of the conduction band. As the Ti-2$p$ states are 
split due to spin-orbit coupling ($2p_{3/2}$ and $2p_{1/2}$ state) the two 
related $L_3$ and $L_2$ edges appear~\cite{stoyanov2007effect}, as denoted in 
Fig.~\ref{fig:Ti-L-ELNES_with_fit}. They can be fitted with five Lorentz 
functions (green dashed lines in Fig.~\ref{fig:Ti-L-ELNES_with_fit}) 
corresponding to different orbitals (relations are given in 
Tab.~\ref{tab:TiO2_energy_splitting}) 
yielding an envelope function (red line in Fig.~\ref{fig:Ti-L-ELNES_with_fit}).

\begin{table}[!ht]
 \centering
 \caption{\label{tab:TiO2_energy_splitting}$d$-orbitals and their
  symmetry label in the setting of the ideal octahedral (O$_{\text{h}}$) 
  and the orthorhombic distorted octahedral (D$_{2\text{h}}$) symmetry 
  group of r-TiO$_2$. The orbitals refer to the local coordinate
  system with the $z$-axis along the orthorhombic distortion.
  The corresponding white lines ($a$, $b$, $b'$, $c$ and $d$) 
  occuring in EEL spectra of the Ti $L_{3\text{,}2}$ edge are given 
  and associated with the energy states.
  }
  \begin{tabular}{l c c c c}
  \hline
  \hline
$d$-orbital 		& symmetry			& symmetry			& $L_3$ white 	& 
$L_2$ white \\
			& label in O$_{\text{h}}$	& label in D$_{2\text{h}}$	
			& line 		& line \\
  \hline
$d_{xz}$  		        & $t_{2g}$			& $b_{2g}$			& $a$		& $c$ \\
$d_{yz}$  		        & $t_{2g}$ 			& $a_{g}$			& $a$		& $c$ \\
$d_{xy}$ 		        & $t_{2g}$ 			& $b_{3g}$			& $a$		& $c$ \\
$d_{3z^{2}-r^2}$ 	& $e_{g}$			& $a_{g}$			& $b$		& $d$ \\
$d_{x^{2}-y^{2}}$ 	& $e_{g}$			& $b_{1g}$			& $b'$		& $d$ \\
  \hline
  \hline
 \end{tabular}
\end{table}

\begin{figure}[!ht]
\centering
\subfloat[]{%
 \includegraphics[clip,width=0.9\columnwidth]{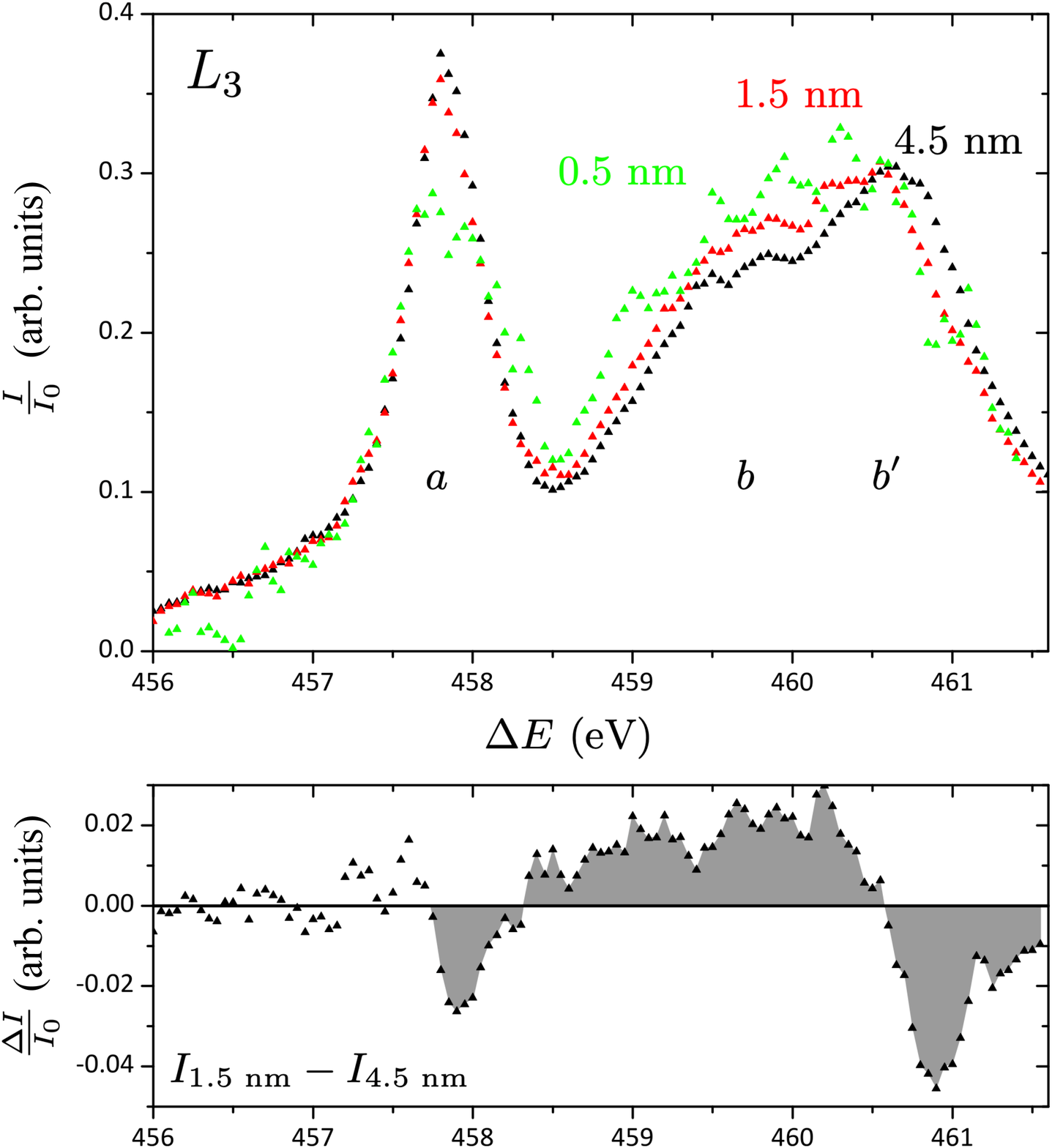}
 \label{fig:EELS_L3_normed}
}

\subfloat[]{%
 \includegraphics[clip,width=0.9\columnwidth]{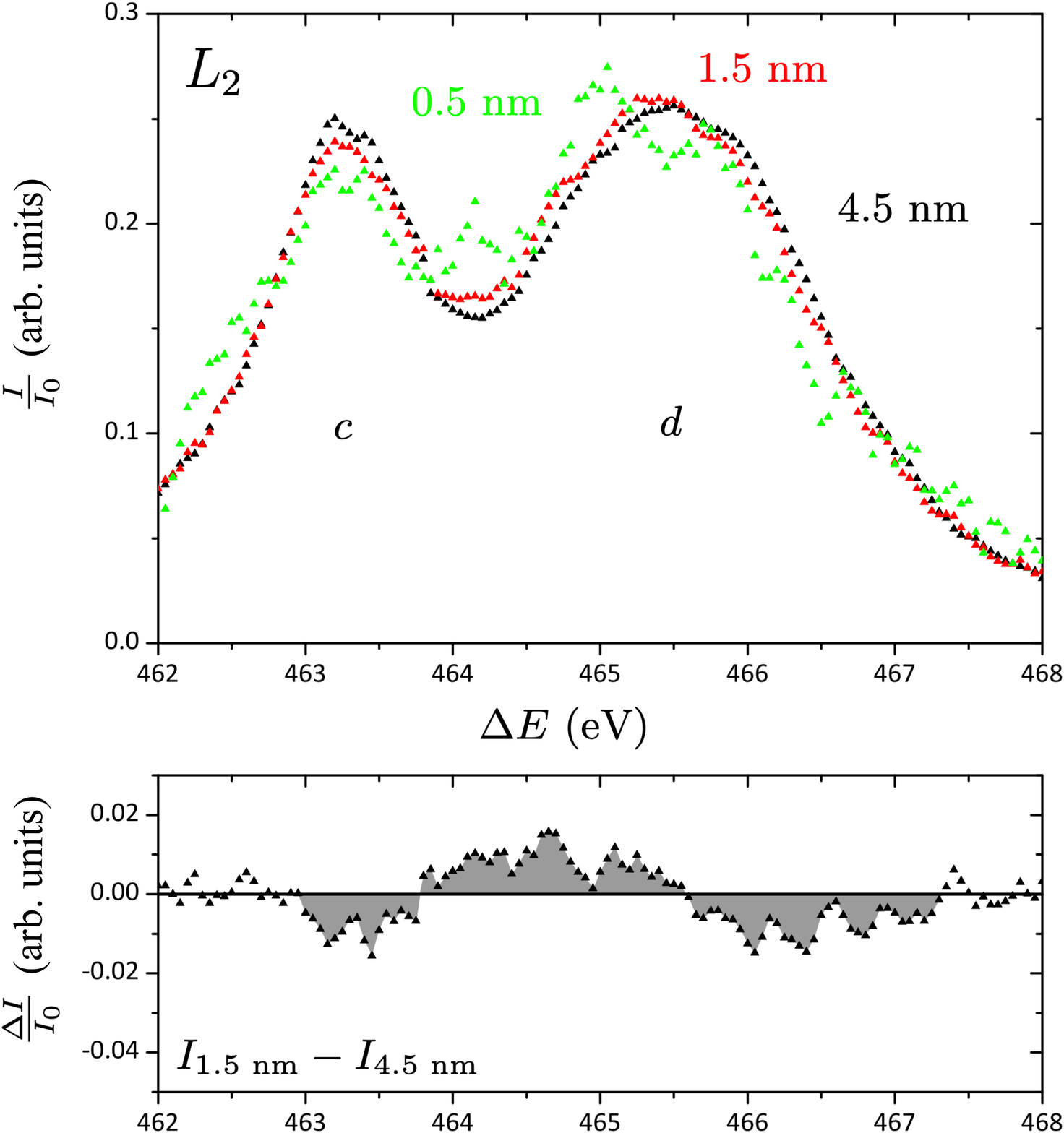}
  \label{fig:EELS_L2_normed}
}
\caption{\label{fig:EEL_spectra}(Color online) Ti $L$ ELNES 
shown for different distances 
to the interface, at a hydrogen partial pressures of $10$~Pa: (a) 
the Ti $L_3$ edge and (b) the Ti $L_2$ edge spectrum in the 
$2$~nm vicinity of the interface. All spectra
are normalized to the area of the respective edge I$_0$. The 
normed intensity difference $\Delta \mathrm{I}/\mathrm{I}_0$ 
between the $1.5$~nm spectrum and the $4.5$~nm spectrum is plotted
in the lower part of each graph. The difference spectra visualize
a shift of states (gray shaded area) in the corresponding Ti $L$ edge.}
\end{figure}

\begin{figure}[!ht]
\subfloat[]{%
 \includegraphics[clip,width=0.9\columnwidth]{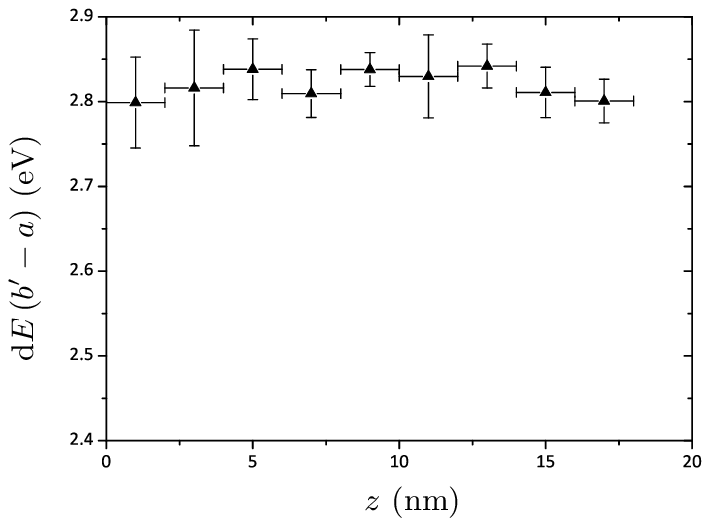}
 \label{fig:energy_splitting_1_Pa}
}

\subfloat[]{%
 \includegraphics[clip,width=0.9\columnwidth]{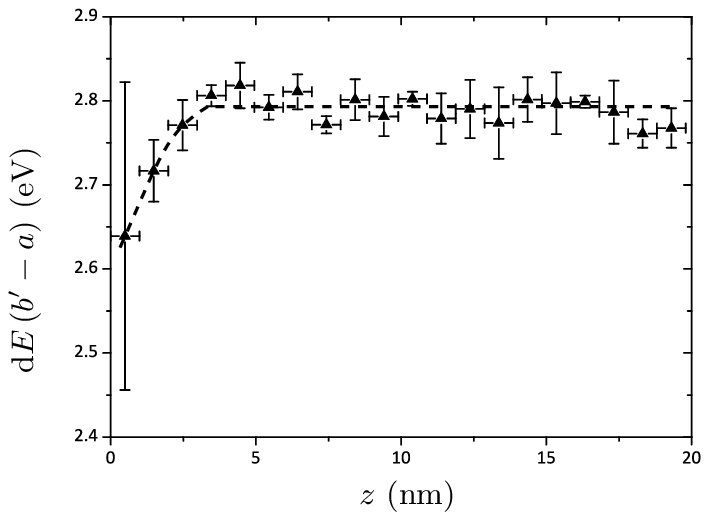}
 \label{fig:energy_splitting_10_Pa}
}
\centering
\caption{\label{fig:energy_splitting}Analysis of Ti $L$ ELNES
on Pd/TiO$_2$, as function of the 
interface distance $z$, for the two selected hydrogen partial pressures (a) 
$1$~Pa and (b) $10$~Pa. Energy splitting d$E(b'-a)$ between the $b'$
and $a$ white lines are shown. (a) an constant energy difference of 
about $2.8$~eV is measured. (b) at $10$~Pa H$_2$ the energy 
splitting decreases near the interface, starting at about $2$~nm 
distance. The black dashed line is plotted to guide the eye.}
\end{figure}

Remarkable differences of the Ti $L$ edge spectrum are observed at a 
hydrogen gas pressure of $10$~Pa, in the vicinity of the interface.
Three representative spectra are shown in
Fig.~\ref{fig:EELS_L3_normed} and \ref{fig:EELS_L2_normed}, for 
different distances from the interface. The Ti $L$ edge
spectrum obtained at $1.5$~nm distance from the interface shows
distinct differences compared to the bulk spectrum, namely the shift by
about $150$~meV in the $b$, $b'$ and $d$ lines region. These differences 
are shown in Fig.~\ref{fig:energy_splitting_10_Pa}. At a distance of 
$4.5$~nm no differences are abserved with respect to the conventional 
bulk spectrum of r-TiO$_2$. Thus, for distances $\geq$ $4.5$~nm
the spectra resemble that of the bulk r-TiO$_2$.
We interpret "extra" features in the spectra close to the interface as 
noise due to the decreasing intensity of the EELS signal. Fitting 
the ENLES with the five Lorentz functions is not influenced by this noise. Therefore 
analysis of the individual peak positions gives, inter alia, 
the energy splitting d$E(b'-a)$.
It should be noted that the method was tested on interfaces before hydrogen loading. 
No changes in the Ti $L$ ELNES could be found here, as function of the interface 
distance.

The energy splitting d$E(b'-a)$ is shown
in Fig.~\ref{fig:energy_splitting} as a function of the distance
from the interface $z$, for hydrogen partial pressures of
$1$~Pa and $10$~Pa. At $1$ Pa H$_2$, d$E(b'-a)$ is about $2.8$~eV and does
not depend on $z$, thereby resembling the
bulk value~\cite{Brydson_1989}. At an H$_2$ partial 
pressure of $10$~Pa, d$E(b'-a)$ decreases by about $150$~meV when 
approaching the interface. This effect sets in below $2$~nm from
the interface. 

The difference spectrum between the near interface ($1.5$~nm) spectrum
and the bulk ($4.5$~nm) spectrum allows visualizing changes in the DoS
and the related orbitals.  These difference spectra are shown in
Fig.~\ref{fig:EELS_L3_normed} and \ref{fig:EELS_L2_normed} below the
corresponding EEL spectra. In the difference spectrum of the $L_3$
edge shown in Fig.~\ref{fig:EELS_L3_normed}, states disappear at about
$458$~eV and $461$~eV (gray shaded area with I/I$_0 < 0$), while new
states appear between $458.35$~eV and $460.55$~eV (gray shaded area
with I/I$_0 > 0$). This indicates that some energy states are shifted
by about $1$~eV to $2$~eV with respect to the EELS in the
bulk r-TiO$_2$. Below about $457.5$~eV the difference signal reflects
the data scatter. The difference spectrum of the $L_2$ edge shown in
the lower part of Fig.~\ref{fig:EELS_L2_normed} shows similar results:
States disappear at about $463.4$~eV and $466.4$~eV, while new states
appear between $463.8$~eV and $465.55$~eV.  The effect in the
difference spectra is stronger for the $L_3$ edge than for the $L_2$
edge. We relate this to the stronger lifetime related broadening in
the $L_2$ edge~\cite{Brydson_1989, Kucheyev_2004}.

Interestingly, the observed small shift of the $b'$ peak by about $150$~meV 
(Fig.~\ref{fig:energy_splitting_10_Pa}) is the result
of a chemically induced shift of states in the order of $1$~eV. If this 
chemically induced shift occurs only on a fraction of Ti sites in the r-TiO$_2$ 
lattice, namely at Ti sites that are affected by a defect, the total EEL 
spectra can show comparably small shifts. Our hypothesis here is, that defects 
are located in r-TiO$_2$ in the close vicinity of the interface of the Pd 
metal. These defects are induced by the presence of hydrogen. The shift of the 
$b'$ peak position, consequently, contains information about the local defect 
concentration. The corresponding defect concentrations are addressed
in Chap.~\ref{chap:thermodynamic_model}.

\subsection{Simulations}
\label{chap:simulation}
In order to shed light on the experimental findings we performed
calculations on hydrogen and oxygen-vacancy related defects. The
nature of the defects is investigated using the projected density of
states, wave function plots and formation energies.

\begin{figure}[!ht]
\includegraphics[width=\columnwidth]{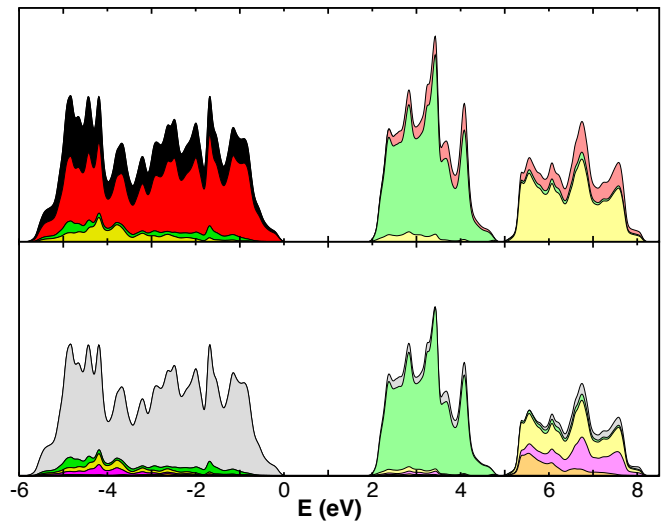}
\centering
\caption{\label{fig:tio2-orb} Density of states (DoS) for r-TiO$_2$.
 The total DoS is given in black and gray, in the top and bottom graph 
 respectively. The projected DoS are
 shown in red for oxygen, in green for $t_{2g}$-like and yellow for
 $e_{g}$-like Ti-$d$ orbitals. Unoccupied states are drawn in a
 lighter color than the filled states. The energy zero is set to
 the top of the valence band. The bottom graph treats the perfect
 oxygen site as a defect with symmetry adapted orbitals.
 Hence, the $A'_1$ (orange), $E'$ (magenta), $\delta$-$e_{g}$ 
 (yellow) and $\delta$-$t_{2g}$ (green) orbitals are shown as
 the sum of the corresponding three Ti neighbors of the oxygen site.
 These symmetry adapted orbitals are defined in 
 Chap.~\ref{sec:computationaldetails}.}
\end{figure}

The calculated DoS of r-TiO$_2$ is shown in Fig.~\ref{fig:tio2-orb}.
The filled valence band, which extends from $-6$ to $0$~eV, is predominantly of 
O-$p$ character (red) with some contribution of Ti-$d$ orbitals (green and 
yellow). They represent the Ti-O bonding states. The Ti-$d$ 
states are located $2-8$~eV above the valence band. From electron counting, 
there is no electron in Ti $d$-orbitals. The approximately octahedral 
crystal-field splitting divides the Ti-$d$ states into nonbonding 
$t_{2g}$-states (light green), between $2-5$ eV and antibonding $e_{g}$-states 
(light yellow), between $5-8$ eV (see Tab.~\ref{tab:TiO2_energy_splitting} for 
the common labeling using the O$_\text{h}$ symmetry).

The bottom graph in Fig.~\ref{fig:tio2-orb} shows the sum of the
three Ti orbitals, which are adjacent to an oxygen site in the perfect
crystal alongside with the total density of states per Ti atom. The Ti 
orbitals are treated in the local D$_{3\text{h}}$ symmetry of a defect 
site as introduced in Chap.~\ref{sec:computationaldetails} and 
Fig.~\ref{fig:orbital-rep}. The $A'_1$ (orange) orbitals are located in the 
bottom part of the $e_{g}$-like states, while the $E'$ 
(magenta) orbitals contribute to the upper part of the $e_{g}$-like states.
 
The calculated structural parameters and heats of formation are given
in Tab.~\ref{tab:tio2-strc}. They are in good agreement with 
experimental results. The band-gap is with
$1.905$~eV for PBE smaller than the experimental value of
$3.05$~eV\cite{pascual78_prb18_5606}. Conventional functionals such as
PBE underestimate the band-gap by approximately $1.15$~eV~\cite{mo95_prb_51_13023}.
As already noted, this finding is well-known for PBE functionals. In the 
following, the various defects are discussed individually.

\begin{table}[!ht]
\caption{\label{tab:tio2-strc}Lattice parameters $a$ and $c$, internal
  parameter $u$~\cite{Cromer_1955} and band-gaps of r-TiO$_2$ 
  from the PBE method, and 
	experiments~\cite{Cromer_1955, finkelstein99_prb60_2212,
  mo95_prb_51_13023}.}
\begin{ruledtabular}
\begin{tabular}{lcc}
Quantity & PBE & Experiment \\
\cline{1-3}
$a$ ({\AA})  		           & 4.562  & 4.593 \\
$c$/$a$                         & 0.644   & 0.644 \\
$u$/$a$                         & 0.302   & 0.305 \\
band-gap (eV)               & 1.905   & 3.05 \\
heat of formation (eV)  &-8.56    &-9.74 \\
\end{tabular}
\end{ruledtabular}
\end{table}

\subsubsection{Interstitial hydrogen}
The PBE calculations on different charge states on the interstitial 
H show that it is only stable in the positive charge state H$_{\text{i}}^+$. 
Thus, the neutral and negative charge states of the interstitial proton 
do not exist. The charge state levels for the transition of the neutral and 
negative charge states lie in the conduction band. This may be due to the 
band gap underestimation of PBE.

\begin{figure*}[htbp]
\includegraphics[width=2.\columnwidth]{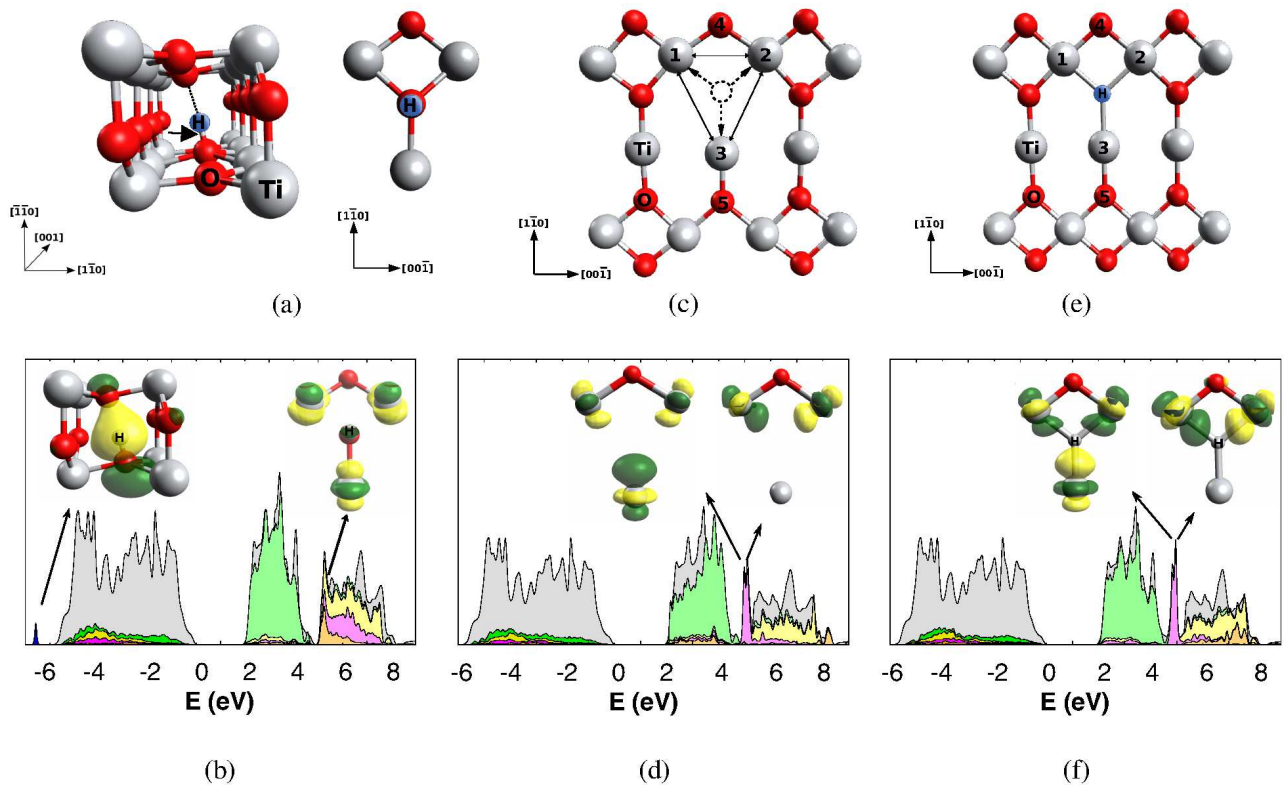}
\caption{\label{fig:tio2-h} Structural and electronic properties of the
  three stable H-related defects in r-TiO$_2$. 
  (a) Illustration of the H position for the positively charged state 
  of the interstitial hydrogen 
  (H$_{\text{i}}^+$). Hydrogen bond is represented by dotted line. 
  (b) Density of states of interstitial hydrogen
  atom H$_{\text{i}}$ in the positive charge state. 
  The projected density of states is colored as
  follows: hydrogen (blue), $A'_1$ (orange), $E'$ (magenta),
  $\delta$-$e_{g}$ (yellow) and $\delta$-$t_{2g}$ (green) orbitals on the
  three Ti neighbors of the defect. To provide context, the
  total density of states of r-TiO$_2$ (grey) is shown as well. The
  projected density of states is related to the colored area, rather
  than the ordinate value. Relevant wave functions are shown as
  isosurface with colors (yellow and green) distinguishing the sign.
  (c) Structure of the oxygen-vacancy (V$_{\text{O}}$) and 
  (d) density of states of the oxygen vacancy
  V$_{\text{O}}$ in the double positive charge state. 
  (e) Structure of the hydrogen-complex H$_{\text{O}}$ with an oxygen 
  vacancy, in r-TiO$_2$ and (f) density of states for the hydrogenated
  oxygen vacancy H$_{\text{O}}$ in the positive charge state.
  }
\end{figure*}

The interstitial hydrogen H$_{\text{i}}^+$ forms an OH bond (as indicated by the arrow 
in Fig.~\ref{fig:tio2-h}a) perpendicular to the Ti-O plane. 
The OH bond length is $1.003$~\AA, about $0.1$~\% smaller than an OH bond 
in water at $60$ K~\cite{kuhs83_jpc87_4312}. The three 
Ti-O bonds close to this O-H bond expand by about $7.2$~\%,
resulting in a trigonal pyramidal distortion of the oxygen
coordination shown in Fig.~\ref{fig:tio2-h}a. The OH group
forms a hydrogen bond (marked by dash line) with one of the oxygen ions on 
the opposite side of the cage. The hydrogen bond of the hydrogen interstitial is 
$1.773$~{\AA} and, therefore, $1.3$~\% larger than that in water with $1.75$~{\AA}. 
The off-axis orientation of the hydrogen bond tilts the OH bond by
$9^\circ$ away from the plane normal of the O-Ti bonds.
The distance of the defect center C$_{\text{D}}$ to the three
neighboring Ti atoms reveals outward breathing by about 
$5.8$~\%. 

The density of states for the interstitial hydrogen H$_{\text{i}}^+$ is shown in 
Fig.~\ref{fig:tio2-h}b using the same color code as described in 
Fig.~\ref{fig:tio2-orb}. 
The total density of states of the supercell per Ti atom (grey) is shown 
alongside with defect symmetry adapted orbitals $A'_1$, $E'$, 
$\delta$-$e_{g}$, and $\delta$-$t_{2g}$. 

The OH $\sigma$-bond state (blue) splits off below the oxygen 2$p$ valence
band at about $-6.5$~eV. The orbital of the $\sigma$-bond is represented in 
an iso-surface plot as depicted in Fig.~\ref{fig:tio2-h}b. 
The state extends along with the hydrogen bond 
to the opposite O-ion. There is no gap state for the interstitial 
hydrogen. Of particular interest concerning the measured EEL spectra are 
the defect induced changes of the Ti-$d$ states forming the conduction
band. In the low energy part of the $e_g$ states at about $5$~eV, we observe 
an accumulation of spectral weight. The orbital representation of this state is 
given as iso-surface plot. We attribute the accumulation of $e_g$ states 
at the bottom of the $e_g$-band to the electrostatic 
attraction of the electrons in the Ti-levels and the nearby positive defect.

\subsubsection{Oxygen vacancy}
In the \{110\} plane of r-TiO$_2$, each O atom has an approximately trigonal 
coordination of three Ti atoms (Fig.~\ref{fig:tio2-h}c, 
highlighted with 1,2 and 3). Thus, the formation of an oxygen vacancy
leaves three Ti ions undercoordinated. The PBE calculations on 
different charge states show that O vacancy is only thermodynamically stable in the 
doubly positive charge state V$_{\text{O}}^{2+}$. Electrons added to this 
state are transferred to the conduction band.

For the oxygen vacancy V$_{\text{O}}^{2+}$, the outward breathing is $12.5$~\%. 
The breathing seems to follow a simple electrostatic argument: The positive 
Ti-ions repel each other.

The DoS for the supercell with the oxygen vacancy V$_{\text{O}}^{2+}$ 
is shown in Fig.~\ref{fig:tio2-h}d. The DoS graph shows that the doubly 
positive charge state does not introduce states into the band-gap. 
The Ti-$d$ states with $A'_1$ character (orange)
are shifted with respect to the undoped material, from the $e_g$-like
continuum downwards to the continuum of the $t_{2g}$-like states and upwards
to the top of the $e_g$-like continuum. The corresponding two states with $E'$ 
character (magenta) are nearly degenerate and form two sharp peaks in the 
quasi gap between the Ti $t_{2g}$- and $e_g$-like states at $5$~eV.

In comparison to the undoped r-TiO$_2$, shown in Fig.~\ref{fig:tio2-orb}, 
a contribution of the $A'_1$ and the 
$E'$ states, which are derived from $e_g$-like states, are now located much 
lower in energy, namely in the $t_{2g}$-like continuum. This results from the
Ti-$d$ orbitals pointing into the vacancy which have lost an antibond with
the former oxygen neighbor. This antibond was responsible for the crystal-field
splitting that shifted these orbitals up, above the $t_{2g}$-like
orbitals in r-TiO$_2$ (Fig.~\ref{fig:tio2-orb}). In the absence of the oxygen 
ion, the $A'_1$ and $E'$ states
are shifted down in energy, so that their energy average lies in the
range of the $t_{2g}$-like bands. The orbital presentations of the two 
situations of the $E'$ states are shown in isosurface plots on top of the DoS in 
Fig.~\ref{fig:tio2-h}d. 

The $E'$ states, in the quasi gap, shift spectral 
weight of the $e_g$ states in the low energy. 

\subsubsection{Hydrogenated oxygen-vacancy} 
After describing the isolated defects, we
investigate here the pair of these two defects, namely the
hydrogenated oxygen vacancy H$_{\text{O}}$. As shown
in Fig.~\ref{fig:tio2-h}e, the hydrogen atom of the H$_{\text{O}}$ is
located almost at the defect center, respectively the position of the
missing oxygen ion. 

The PBE calculations on different charge states on the hydrogenated oxygen 
vacancy show that it is thermodynamically stable only in the positive charge 
state H$_{\text{O}}^+$.

The corresponding DoS for the hydrogenated oxygen vacancy is shown in 
Fig.~\ref{fig:tio2-h}f. The DoS graph resembles that of 
Fig.~\ref{fig:tio2-h}d for the V$_{\text{O}}^{2+}$. In 
comparison to the oxygen vacancy, the
dominant effect is to shift the defect level of $A'_1$
character (orange) of the oxygen vacancy downward into the
continuum of the valence band. The hydrogenated
oxygen vacancy H$_{\text{O}}^+$ also does not have states in the band-gap.

The Ti $d$-states with $E'$ symmetry (magenta) exhibit sharp peaks and remain
in the quasi gap between the $t_{2g}$- and the $e_{g}$-like states,
but centered in it. With respect to the oxygen vacancy, they are slightly 
shifted to lower energies. Hence, regarding
these states, the hydrogenated oxygen vacancy is very similar to the bare
oxygen vacancy. We interpret this again with the loss of the oxygen 
antibond. With respect to the EELS measurements, these quasi gap states
can shift spectra weight. The orbital represemtation of the $E'$ states are shown 
in the iso-surface plots on top the DoS in Fig.~\ref{fig:tio2-h}d.

Based on the simulation results, we note that all of the defects deliver a qualitative 
explanation for a shift of states in the unoccupied part of the DoS which can 
explain the EEL measurements.

\subsection{Defect electrochemistry of the interface and the bulk}
\label{chap:thermodynamic_model}
To get a deeper understanding about which of the defects are responsible for 
the EELS changes, we combine the DFT results with statistical
mechanics to simulate the Pd/r-TiO$_2$ interface.

The total energies from the DFT calculations allow one to estimate the
defect concentrations from a thermodynamic multistate defect (TMD)
model.  Furthermore, it allows one to link the concentration to the
spatial position in the sample.  The TMD model is described in
Appendix~\ref{sec:model}. It is based on a multistate description for
the oxygen-lattice sites. Each state $X_\sigma$ corresponds to a
specific defect type and charge state. The defects states used in the
present study are specified in table~\ref{tab:pottstate}, along with
their energy $E_{\sigma}$ and excess number of electrons, hydrogen and
oxygen atoms.  The TMD model accounts for the energy and the
configurational entropy of the defect distributions.  In addition,
electrons and holes are considered within the effective-mass
approximation (see Eq.~\ref{eq:dedv}). The electrostatic interaction
between charge carriers and defects is described as mean field
obtained from the Poisson equation (see Eq.~\ref{eq:poisson}).

The purpose of these model calculations is to elucidate three
questions which emerge from the experimental results presented 
in section~\ref{sec:eels-exp}:
\begin{itemize}
\item Does the potential step between the interface and the bulk
  affect the defect concentrations sufficiently to explain the
  observed spatial profile of the EELS?
\item Is there a relevant increase in the defect concentration by
pressure increase from 1 to 10 Pa?
\item Does the interface concentration of defects increase up to
  a measurable concentration of the order of one per oxygen site in 
  the pressure range of 10 Pa? 
\end{itemize}

\begin{table}[!htb]
\caption{\label{tab:pottstate}Defect states $X_\sigma$ of an oxygen
  site, their energies $E_\sigma$ and excess particle numbers 
  (stoichiometric factors) $\eta_{e,\sigma}$ for electrons, 
  $\eta_{H,\sigma}$ for hydrogen atoms, and $\eta_{O,\sigma}$ oxygen 
  atoms. The excess particle numbers are counted relative to an ideal bulk
  site.}
\begin{center}
\begin{tabular}{c c c r r r}
\hline
\hline
$\sigma$ & $X_\sigma$ &
$E_\sigma-\sum_{k\in{\{e,O,H\}}}\eta_{k,\sigma}\mu_k^{ref}$& 
$\eta_{e,\sigma}$ & $\eta_{H,\sigma}$ & $\eta_{O,\sigma}$ \\
\hline
$1$ & bulk  & $0.000$~eV & $0$ & $0$ & $0$ \\
\hline
$2$ & H$_{\mathrm{i}}^{+}$  & $-1.59018$~eV & $-1$ & $+1$ & $0$ \\
\hline
$3$ & V$_{\mathrm{O}}^{2+}$ & $0.78779$~eV & $-2$ & $0$ & $-1$ \\
\hline
$4$ & H$_{\mathrm{O}}^{+}$  & $1.90610$~eV & $-1$ & $+1$ & $-1$ \\
\hline
\hline
\end{tabular}
\end{center}
\end{table}

In the following, we describe the processes, which emerge from our
model. The treatment considers the experimental conditions during
sample preparation as well as before and after hydrogen
exposure. 

\subsubsection{Bulk oxide before hydrogen exposure}
\label{chap:thermodynamic_model_bulk_rutile}

\begin{figure}[!htbp]
\begin{center}
\includegraphics[width=1.0\linewidth,clip=true]{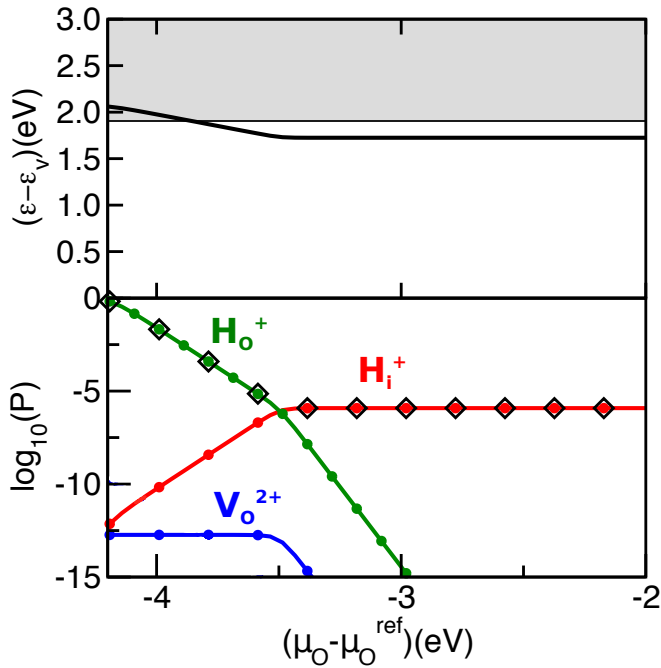}
\end{center}
\caption{\label{fig:oxideofmuo}Calculated Fermi level $\mu_e$ relative
  to the valence band top $\epsilon_v$ (top) as well as defect and
  charge carrier concentrations (bottom) in bulk r-TiO$_2$ as function
  of the oxygen chemical potential $\mu_{\text{O}}$.  Data are shown
  for room temperature. The atmospheric hydrogen partial pressure is
  taken into account.  The concentrations in terms of probability $P$
  per O site of interstitial hydrogen H$_{\text{i}}^+$ are shown in
  red, those of oxygen vacancies V$_{\text{O}}^{2+}$ are shown in blue
  and those of hydrogenated oxygen vacancies H$_{\text{O}}^+$ are
  shown in green.  The concentration of electrons in the conduction
  band is represented by the open diamonds. The shaded region of the
  top graph indicates the difference between the experimental and the
  calculated band-gaps. The latter are obtained from the Kohn-Sham
  levels.  }
\end{figure}

Fig.~\ref{fig:oxideofmuo} shows the concentrations of defects and
electrons in bulk r-TiO$_2$ as a function of the oxygen chemical
potential $\mu_{\text{O}}$ at room temperature and ambient
pressure. Before the Pd film is deposited, the oxygen chemical
potential $\mu_{\text{O}}$ is controlled by the oxygen partial
pressure and the temperature of the ambient.

The natural atmospheric hydrogen content of $0.5$~ppm is taken into
account~\cite{Klose_2015}.  Despite the small hydrogen content of the
atmosphere, its partial pressure is sufficiently high to dominate the
defect chemistry of r-TiO$_2$ at room temperature. Hydrogen related
defects dominate for both, oxygen-poor and oxygen-rich conditions.

Fig.~\ref{fig:oxideofmuo} also summarizes the Fermi level position
(top) and charge carrier concentrations (bottom) in bulk r-TiO$_2$ as
a function of the oxygen chemical potential $\mu_{\text{O}}$. The bulk
Fermi level is determined by the charge neutrality condition. The
Fermi level is measured relative to the valence-band top $\epsilon_v$
and the oxygen chemical potential is measured relative to reference
$\mu_{\text{O}}^{ref}:=\frac{1}{2}E[\text{O}_2]$.  

Under oxygen-rich conditions, that is above
$\mu_{\text{O}}=\mu_{\text{O}}^{ref}-3.5$~eV, the dominant defect is
the hydrogen interstitial H$_{\text{i}}^{{+}}$ as shown in
Fig.~\ref{fig:oxideofmuo}. Our model estimates an H$_{\text{i}}^{{+}}$
concentration (red in Fig.~\ref{fig:oxideofmuo}) of about
\gga{$10^{-6}$} per oxygen site or $\gga{10^{17}}$~cm$^{-3}$. The
positive charge of the defects is compensated by the conduction
electrons.  For this oxygen-rich region, the Fermi level lies
$\gga{0.18}$~eV below the conduction band edge.

Under oxygen-poor conditions, that is for effective chemical
potentials below $\mu_{\text{O}}=\mu_{\text{O}}^{ref}-\gga{3.5}$~eV,
the positive hydrogenated oxygen vacancy H$_{\text{O}}^+$ (green in
Fig.~\ref{fig:oxideofmuo}) is the dominant defect. The concentration
of hydrogenated oxygen vacancies rises strongly with decreasing oxygen
chemical potential. To maintain charge neutrality, also the
conduction-electron concentration increases alongside with the defect
concentration as marked with open black diamonds. The increasing conduction
electron density is reflected by the increasing Fermi level as shown
in the top of Fig.~\ref{fig:oxideofmuo}. 

Below $\mu_{\text{O}}-\mu_{\text{O}}^{ref}=-\gga{4.2}$~eV, our model
predicts defect concentrations of the order of one per oxygen site
(green curve in Fig.~\ref{fig:oxideofmuo}), which marks the limit of
its applicability. Such large defect concentrations indicate that the
material may be close to a phase transition.

The effective oxygen chemical potential in the material is strongly
preparation dependent because the oxygen exchange with the atmosphere
is effectively suppressed at room temperature. The exchange requires
firstly sufficiently large defect mobility, and secondly an
effective surface reaction such as the oxidation of oxygen vacancies
at the surface.  Both are not present at room
temperature~\cite{Haul_1965, Arita_1979, Derry_1981}. Therefore, the material
memorizes the defect concentration of the last high-temperature
preparation step, which allowed the oxide to exchange oxygen
effectively with the surrounding atmosphere.

In the following, we estimate the defect concentrations originating from the experimental 
treatments. In our experiment, the r-TiO$_2$ crystals were first annealed at 
1173~K at approximately $10^{5}$~Pa O$_2$. Subsequently, they have been cooled 
in the same oxygen atmosphere. Then, they were kept under ambient conditions 
for several hours or days before the Pd films were deposited at 1023~K and at 
O$_2$ partial pressure of $7\times10^{-3}$~Pa. The samples remained in this
environment for half an hour before they were cooled down in the same
atmosphere. The material is then stored under ambient conditions. After
preparation, the TEM lamella was transferred to the ETEM,
where it has been kept for about 12 h in vacuum conditions of
$10^{-4}$~Pa. The sample was then exposed to a fixed hydrogen partial
pressure of 1 or 10~Pa of H$_2$. After one hour of hydrogen exposure, 
the EELS measurement was carried out.

\begin{table}[!htb]
\caption{\label{tab:chempots}Partial pressures and chemical potentials
  for (1) ambient conditions, (2) high-temperature anneal under oxygen
  atmosphere, (3) Pd deposition in vacuum, (4) ambient conditions
  after Pd deposition with frozen vacancy concentration V$_\text{O}$ and
  H$_\text{O}$. (5) hydrogen exposure 1~Pa, (6) hydrogen exposure 10~Pa. The
  chemical potentials are specified as relative values
  $\Delta\mu_{\text{O}}=\mu_{\text{O}}-\mu_{\text{O}}^{ref}$ and
  $\Delta\mu_{\text{H}}=\mu_{\text{H}}-\mu_{\text{H}}^{ref}$. Nrs. 4, 5, 6,
  show the equivalent partial pressures, after the Pd film prohibits
  oxygen exchange with the atmosphere.}
\begin{center}
\begin{tabular}{c c c c c c}
\hline
\hline
Nr. & T (K) & $p_{\mathrm{O}_2}$ (Pa)  & $p_{\mathrm{H}_2}$ (Pa) 
& $\Delta\mu_\mathrm{O}$ (eV)  & $\Delta\mu_\mathrm{H}$ (eV) \\
\hline
1 &  293 & $10^{+5}$     & $5\times 10^{-2}$ & $-0.218$ & $-0.210$ \\
2 & 1173 & $10^{+5}$          & 0 & $-1.271$ & $-\infty$  \\
3 & 1023 & $7\times10^{-3}$ & 0 & $-1.806$ & $-\infty$ \\
4 &  293 & $0.3\times10^{-107}$ & $5\times 10^{-2}$  & $-3.49$ & $-0.210$ \\
5 &  293 & $2\times10^{-107}$ & 1   & $-3.47$ & $-0.172$  \\
6 &  293 & $6\times10^{-107}$ & 10  & $-3.45$ & $-0.143$  \\
\hline
\hline
\end{tabular}
\end{center}
\end{table}

In Tab.~\ref{tab:chempots}, the experimental conditions are translated
into oxygen chemical potentials as described in
Appendix~\ref{app:partpress}.  We assume that thermodynamic
equilibrium can be reached under the chosen annealing conditions,
while the Pd film prevents oxygen exchange with the atmosphere.

As a reference, we list the partial pressures and chemical potentials
under ambient conditions.  The hydrogen chemical potential was
determined by using the natural content of $0.5$~ppm of hydrogen in
the air.

The oxygen content of the oxide, i.e. its oxygen vacancy
concentration, is determined in the high-temperature processing steps.
Our thermodynamic model predicts via Eq.~\ref{eq:probability}
a concentration of $2\times 10^{-7}$ oxygen vacancies per site for
the 1173~K annealing at $10^5$~Pa and of $7\times10^{-7}$
vacancies per site for the Pd deposition at 1023~K in a vacuum.  

We assume that the oxygen-vacancy concentration is
effectively maintained after the Pd deposition at high-temperatures,
because the Pd film prevents diffusion of oxygen from the atmosphere
into the sample.  Therefore, we assume a remaining oxygen vacancy
concentration, hydrogenated or not, of $7\times 10^{-7}$ per oxygen
site after sample preparation.

After deposition of the Pd film, the total concentration of oxygen
vacancies, hydrogenated and not, is frozen in.  Therefore, while the
material cools down to room temperature, the oxygen chemical potential
moves downward. At room temperature, the concentration of $7\times
10^{-7}$ oxygen vacancies, present during Pd deposition, translates
into an oxygen chemical potential of
$\mu_{\text{O}}=\mu_{\text{O}}^{ref}-\gga{3.5}$~eV. For a
  hypothetical environment that can exchange oxygen with the probe,
  the oxygen chemical potential, respectively the oxygen vacancy
  concentration, translates into an ultra-low equivalent partial
  pressure of $10^{-107}$~Pa.

  As seen in Fig.~\ref{fig:oxideofmuo}, after the sample preparation
  the material is thus at the boundary
  $\mu_{\text{O}}=\mu_{\text{O}}^{ref}-\gga{3.5}$~eV between
  electron-rich conditions, dominated by interstitial hydrogen, and
  electron-poor conditions, dominated by hydrogenated oxygen
  vacancies.

The Pd film is further important as a catalyst for hydrogen
dissociation~\cite{Wicke_1978}. As the thin TEM lamella is exposed to
air, hydrogen can diffuse into it and saturate the existing oxygen
vacancies, which results in the defect concentrations presented in
Fig.~\ref{fig:oxideofmuo}.

\subsubsection{Defect profile at the metal-oxide interface}
\label{chap:thermodynamic_model_interface}
Above we considered the defect concentrations in the bulk, where they
are governed by local charge neutrality. In the following, we discuss
the interface and the influence on hydrogen loading.

While the analysis predicts an equal concentration of hydrogenated
oxygen vacancies H$_{\text{O}}^+$ and interstitial hydrogen H$_{\text{i}}^+$,
interstitial hydrogen is the dominant defect under hydrogen exposure.
This allows us to simplify the following discussion by limiting it to
a single defect, namely H$_{\text{i}}^+$, albeit the calculations have been
performed with the complete model.

With the above-mentioned limitation, the H$_{\text{i}}^{+}$ defect
concentration per oxygen site, $P_{\text{H}_{\text{i}}^{{+}}}$, can be
expressed in terms of the hydrogen chemical potential $\mu_\text{H}$ and the
Fermi level $\mu_e$ (Eq.~\ref{eq:probability}),
\begin{eqnarray}
P_{\text{H}_{\text{i}}^{{+}}}&=\Bigl[1+{\rm e}^{\beta(E[\text{H}_{\text{i}}^{{+}}]
-E[B]+(\mu_e-v_e(\vec{r}))-\mu_\text{H}})\Bigr]^{-1}\;,
\label{eq:hiplusdefectprobability}
\end{eqnarray}
where $v_e(\vec{r})=-e\Phi(\vec{r})$ is the potential acting on the
electrons due to the electric potential $\Phi(\vec{r})$. Per
convention, $v_e(\vec{r})$ is equal to the energy of the local
valence-band top.

The energy $\mu_e-v_e(\vec{r})$ is required to lift the electron to the
Fermi level, when the charged defect H$_{\text{i}}^+$ is formed from
the neutral H$_{\text{i}}^{(0)}$. This energy differs between the bulk
and the interface and, therefore, also the concentrations are
different. 

Increasing the hydrogen partial pressure of the environment enhances
the bulk concentration of H$_{\text{i}}^+$. The charge neutrality
condition pushes the Fermi level upward, which in turn increases the
number of electrons in the conduction band. The high position of the
Fermi level counteracts the formation of additional charged defects
until charge neutrality is reached.

At the interface, this mechanism works entirely different. Rather than
lifting the electron towards a Fermi-level near the conduction band
edge, it can be placed into the metal.  At the interface, the Fermi
level is pinned by the Schottky barrier at a specific energy
$\Phi_{Bn}$ below to the local conduction band of the oxide. This
energy $\mu_e-v_e(\vec{r}_I)=\epsilon_c-\Phi_{Bn}$ lies further below
the local conduction band edge $\epsilon_c+v_e(\vec{r})$ than in the
bulk, which stabilises positively charged defects near the interface.
For sufficiently low defect concentrations, the enhancement
of defects at the interface can be related to the band bending
$\Delta{v}_e=v_e(z=0)-v_e(z=\infty)$, where $z$ is the distance from
the interface, as $\e{\beta\Delta{v}_e}$. Approximating the band
bending by the Schottky barrier yields an interface enhancement of
the defect concentration by a factor $10^{13}$ relative to that of
the bulk oxide.

\subsubsection{Hydrogen loading}

Hydrogen related changes of the EEL spectrum are observed in
the experiment for a concentration of hydrogen related defects per
oxygen site in the order of 10~\%, as estimated from the shift of 0.15~eV
in the measured EELS (Fig.~\ref{fig:energy_splitting}b) and the
defect induced shift of calculated energy levels in the range
1-2~eV. (Fig.~\ref{fig:tio2-h})

The crossover to a hydrogen concentration of order $1$ is estimated
from Eq.~\ref{eq:hiplusdefectprobability} as the hydrogen chemical
potential $\mu_\text{H}=E[H_i^{+}]-E[B]+\mu_e-v_e(\vec{r})$.  At the
interface this value is reached for $\mu_H^*=\mu_{H}^{ref}-0.47$~eV,
whereas in the bulk it is reached substantially later, namely at
$\mu_H^*=\mu_{H}^{ref}+0.31$~eV. This reflects the large difference of
hydrogen related defects at the interface and in the bulk.
Experimentally, this crossover takes place above a hydrogen chemical
potential of $\mu_{\text{H}}=\mu_{\text{H}}^{ref}-0.143$~eV,
corresponding to a hydrogen partial pressure of 10~Pa.

There is a discrepancy between the experimentally observed crossover
at $\mu_H^*>\mu_{H}^{ref}-0.143$~eV (see Table~\ref{tab:chempots}) and
the prediction from the TMD model of
$\mu_H^*=\mu_{H}^{ref}-0.47$~eV. We attribute this discrepancy of
$0.33$~eV to the uncertainties of the parameters used in the TMD
model. The most likely source of error is the underestimation of the
band gap by 1.1~eV in DFT.  This underestimates the energy $E_\sigma$
of positive defects.  Furthermore, the band gap underestimation hides
defect-induced charge state levels below the conduction band
edge. These charge state levels may pin the Fermi level and thus limit
the concentration of positively charged defects.  This in turn, would
reduce the band bending and consequently the interface enhancement of
positively charged defects. A second possible source of errors is the
value of the Schottky barrier, which has not been directly measured
but deduced from the Schottky model. In view of these uncertainties,
we attribute the deviation of $\mu_H^*$ between theory and experiment
to the limitations of the theory.

\begin{figure}[!ht]
 \includegraphics[width=0.9\linewidth,clip=true]{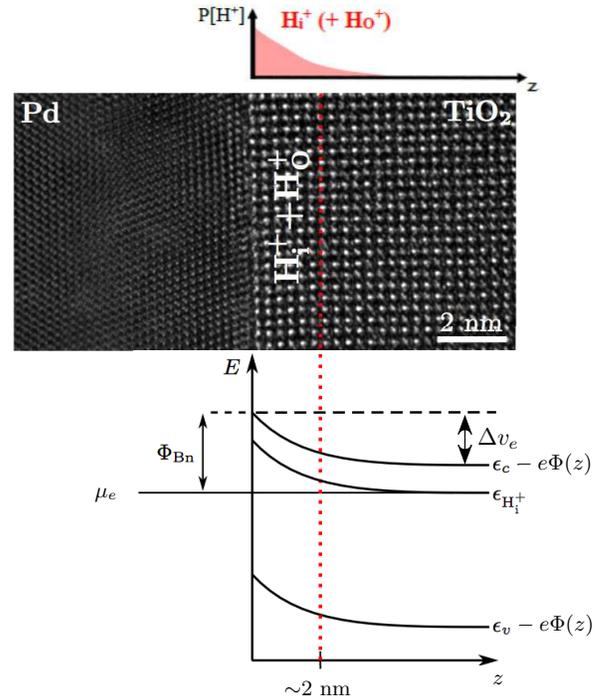}
 \caption{\label{fig:Pd-TiO2_with_H} Schematic diagram of the
  electronic structure of the Pd(111)/TiO$_2$(110) close to the 
  interface in hydrogen environment. Due to the presence of the
  Schottky barrier $\Phi_{\text{Bn}}$, hydrogen-related defects
  enrich at the interface and its vicinity. Mainly the concentration
  of interstitial hydrogen H$_{\text{i}}$ is influenced by the 
  applied hydrogen partial pressure within the ETEM.
  }
\end{figure}

Based on our findings, the inhomogeneous defect distribution in $z$ is the 
effect of positively charged mobile hydrogen species. From these findings, we 
derived a schematic band picture as shown in Fig.~\ref{fig:Pd-TiO2_with_H}. The 
corresponding hydrogen-related defects are present in the $2$~nm vicinity of the 
Pd/TiO$_2$ interface.

\section{Summary and Conclusions}
In a combined study using ELNES, ETEM, and DFT we investigated the
Pd/r-TiO$_2$ interface under various hydrogen partial pressures. The
energy splitting between $b'$ and $a$ bands in the Ti
$L_{3}$ ELNES was found to be decreased for hydrogen partial pressures
exceeding 10~Pa. This effect is limited to 2~nm from the
interface. 

The calculated DoS of defects (H$_{\text{i}}^+$, V$_{\text{O}}^{2+}$,
and H$_{\text{O}}^+$) qualitatively explain the origin of these
hydrogen-induced changes of the ELNES.  The shift or the anti-bonding
Ti-$d$ orbitals is attributed to the Coulomb potential from the positive
defect located at an adjacent oxygen site.

The defect distribution has been studied with a TMD model using energies from 
DFT calculations. The TMD model predicts the spatial distribution of defects  in 
thermal equilibrium and the resulting electrostatic potential near a 
metal-semiconductor interface. Thus, it generalizes the classical 
metal-semiconductor theory which studies stationary 
dopants~\cite{moench2004electronic}.

The TMD model predicts that, despite the small hydrogen concentration below the 
ppm range in the atmosphere, hydrogen-related defects (H$_{\text{i}}^+$ and 
H$_{\text{O}}^+$) are the dominant defects in bulk TiO$_2$. We expect this to be 
valid also for a broader class of materials. The defect H$_{\text{O}}^{+}$ 
dominates in oxygen-poor conditions, while the interstitial hydrogen 
H$_{\text{i}}^{+}$ is the dominant defect under oxygen rich conditions. 

An analysis of the processing steps with the TMD model estimates that
our samples are at the transition from the oxygen poor regime, where
the hydrogenated oxygen vacancy (substitutional hydrogen)
H$_{\text{O}}^{+}$ dominates, and the oxygen-rich regime, where
interstitial hydrogen H$_{\text{i}}^{+}$ is the dominant defect.
This implies that interstitial hydrogen
H$_{\text{i}}^{+}$ becomes the dominant defect, when the hydrogen
partial pressure is increased.

According to the TMD model, the concentration of positively charged
defects is enhanced at the interface compared to the bulk oxide.  This
increases, in particular, the interface concentration of the dominant
hydrogen related defect, namely H$_{\text{i}}^+$.

Increasing the hydrogen partial pressure to 10~Pa, as in our EELS
experiments within the ETEM, increases the interstitial hydrogen
defect concentration further. Changes in the EELS are expected to
become visible at an interfacial defect concentration in the range of
10~\%. Such a defect concentration is compatible with our
calculations, even though the required hydrogen chemical potential is
predicted 0.33~eV lower than observed in experiment.

The defect density in r-TiO$_2$ is a decaying distribution
of positively charged mobile hydrogen species starting from
interface to the bulk.  The presence of the defect states can
explain an increased conductivity of the metal/TiO$_2$ contacts in
hydrogen
gas~\cite{kobayashi1994reactions,cerchez2013dynamics,strungaru2015interdependence}.
Strungaru \textit{et al.}~\cite{strungaru2015interdependence} explain
gas sensing measurements on platinum/TiO$_2$ contacts with the
presence of interstitial hydrogen and hydrogen-oxygen vacancy
complexes. Based on the presented results we suggest that interstitial
hydrogen is responsible for the changed electrical properties of the
metal/semiconductor contact.

The present study offers an example of hydrogen interacting with 
the Pd/r-TiO$_2$ model system. Hydrogen induced changes in the close 
vicinity of the Pd/r-TiO$_2$ interface, give important insights 
for understanding gas sensing and catalytic processes of buried interfaces. 
The described change of the mobile defect concentration at metal/oxide 
interfaces also concerns the perimeter of catalytically active metal particles 
supported by an oxide or vice versa. With this paper, we open a door to a new 
view on the physics at interfaces, that include the behavior of mobile defect 
species.

\section*{Acknowledgements}
We thank fruitful discussions with Thorsten Stolper and Felix Jung.
We acknowledge the help of Niklas Herwig for AFM and XPS, as well as 
Julius Scholz for XPS measurements. The use of equipment in the Collaborative 
Laboratory and User Facility for Electron Microscopy (CLUE) is gratefully 
acknowledged. Financial supports from the CRC 1073 (projects 
B03, C03, C04, C06, and Z02), PU131/9-2, and 
Project ID 390874152 (POLiS Cluster of Excellence) of the Deutsche 
Forschungsgemeinschaft (DFG) are gratefully acknowledged. J.C. acknowledges 
financial support from the Czech Science Foundation (project 21-16218S).

\appendix
\section{Variable energy positron annihilation spectroscopy (VEPAS)}
\label{chap:appex_VEPAS}
\begin{figure}[!ht]
\includegraphics[width=1.0\columnwidth]{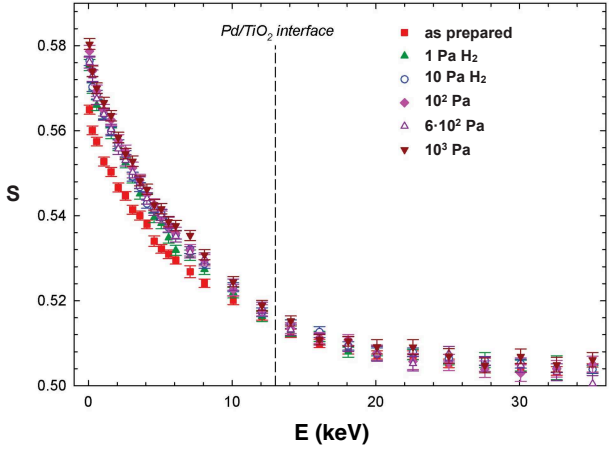}
\centering
\caption{VEPAS on the Pd/r-TiO$_2$ system shows the $S$ parameter
as a function of the positions incident energy $E$. No changes of
the $S$ parameter are found at the interface or in the r-TiO$_2$
bulk.
}
\label{fig:VEPAS}
\end{figure}

Variable energy positron annihilation spectroscopy 
(VEPAS)~\cite{schultz1988interaction} was employed to complementary 
characterize the defect depth profile in hydrogen gas exposed Pd/r-TiO$_2$ samples.
The results are shown in Fig.~\ref{fig:VEPAS}. 
The VEPAS studies were carried out on a magnetically guided slow 
positron beam~\cite{Anwand_2012}. The energy of incident 
positrons was varied in the range from $80$~eV up to $35$~keV. The doppler 
broadening of the annihilation photopeak was measured by a HPGe 
detector with the energy resolution of $1.09$~keV at $511$~keV. The shape of 
the Doppler broadened annihilation photopeak was characterized using the $S$ 
(sharpness) parameter~\cite{krause1999positron}. At very low incident 
energies ($E < 1$~keV) positrons annihilate almost exclusively on 
the sample surface. With increasing energy, positrons penetrate deeper 
and deeper into the Pd over-layer and the fraction of positrons 
diffusing back to the surface decreases. A further increase of the
incident energy ($E > 13$~keV) allows positrons to penetrate into 
the r-TiO$_2$ region. Finally, at high incident energies ($E > 30$~keV)
virtually all positrons are annihilated in r-TiO$_2$ and $S$ 
reaches a bulk value. No changes of the bulk $S$ value caused by
hydrogen loading (up to H$_2$ pressure of $10$~mbar) were observed
at high energies. Similarly, no changes of $S$ were found at the
Pd/TiO$_2$ interface at a positron energy of $E \approx 13$~keV
corresponding to the mean penetration depth of $200$~nm. This
testifies that hydrogen loading did not introduce any defects 
capable of positron trapping neither into the r-TiO$_2$ bulk nor 
into the Pd/r-TiO$_2$ interface. Note that only electrically neutral or 
negatively charged defects are capable of positron trapping while 
positively charged defects repel positrons and cannot be observed by 
VEPAS. In addition, enough open volume is needed for positron trapping. 
Oxygen vacancies are too small to trap positrons, whereas Ti vacancies 
yield enough open volume. This was investigated by positron density 
simulations in r-TiO$_2$~\cite{Bongers_2018}.
Hence, from the VEPAS results one can conclude that hydrogen 
loading did not introduce Ti vacancies but no conclusion about oxygen
vacancies can be made. However, changes of the $S$ parameter are
found in the Pd over-layer (E $<$ 13~keV). This effect is discussed
by Roddatis \textit{et al.}~\cite{Roddatis_2018}.

\section{Advanced EELS procedure}
\label{chap:appex_EELS_procedure}
\begin{figure}[!ht]
 \centering
\includegraphics[width=0.8\columnwidth]{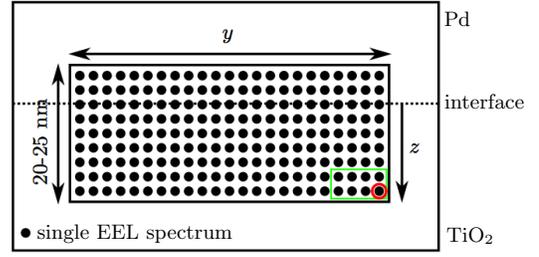}
 \caption{\label{fig:EELS_experiment}Scheme of the rectangular 
 SI pattern in the Pd/TiO$_2$ interface region. The dots 
 represent the positions where EEL spectra are recorded 
 (beam diameter $\approx 5$~\AA{}). Their number depends on the
 chosen size of the SI pattern and the used step size. $z$ defines
 the distance to the interface. $y$ varies between $20$~nm and 
 $40$~nm. The red circle indicates the start position of the
 measurement. It first moves in negative $z$ direction and than
 continues in the same way in the next row to the left. The green
 box indicates the spectra that are binned afterwards to improve
 the signal-to-noise ratio.}
\end{figure}

For the EELS studies at the interface, a procedure was necessary that combines high energy 
resolution, low energy drift and a good signal to noise ratio.
In contrast to other suggested methods~\cite{Heidelmann_2009,Sader_2010}, 
the procedure presented here offers 
error bars on electron energy-loss (EEL) spectra.
Rectangular spectrum imaging (SI) pattern are chosen to measure a
whole set of EEL spectra on the flat and sharp interfaces.
This pattern is indicated schematically in 
Fig.~\ref{fig:EELS_experiment}. The measurement protocol 
describes as follows:
A SI pattern is placed on the region of interest which is, here, at the 
Pd/TiO$_2$ interface, see Fig.~\ref{fig:EELS_experiment}. Thus, 
the corresponding EEL spectra are collected at defined positions 
which are marked with red dots in Fig.~\ref{fig:EELS_experiment}. 
The electron beam starts scanning at the red marked position 
and continues in negative $z$ direction. The $z$ coordinate is 
defined as the distance to the Pd/TiO$_2$
interface. This interface ($z=0$) is here chosen as that location where 
the EEL spectrum does not 
contain any significant Ti $L$ edge signal. Afterwards, the measurement 
continues in the same way in the next row to the left. 
Vertical alignment of highest interface sharpness was provided by 
studying a tilting series. The lateral interface homogeneity was verified by 
repeated measurements at different interface positions. Assuming lateral 
interface homogeneity, the important changes are here expected just to 
depend on $z$. This allows binning the acquired EEL spectra (as indicated 
with the green box) to enhance the signal-to-noise ratio (S/N).

Details on the measurements and data evaluation is described in the following.
A more detailed description is given by Bongers~\cite{Bongers_2018}.
\begin{enumerate}
  \item The Ti $L$ edge is studied by EELS in monochromated scanning
    TEM (STEM) mode (energy resolution $200$~meV, probe size $\approx 5$~\AA)
    as a function of the distance to the interface $z$ and of the H$_2$ 
    pressure. The final energy resolution of the EEL spectra in the 
    r-TiO$_2$ was better than $500$~meV. 
    beam currents and integration times are about $130$~pA and about
    $50$~ms per EEL spectrum. The choice of a relatively low integration
    time (seconds can easily be achieved) for the core loss region 
    results in low energy drift of the total spectrum and allows for a 
    good energy alignment with Dual EELS. The low loss region is used for 
    this alignment. The chosen dispersions for Dual EELS are $0.05$~eV/ch 
    and $0.1$~eV/ch. A rectangular SI pattern (compare
    Fig.~\ref{fig:EELS_experiment}), aligned perpendicular to the 
    interface, is used to record the EEL spectra with step sizes of 
    $0.5$~nm and $1$~nm. The total collection time for one SI pattern 
    is about $60$~s to minimize spatial sample drift effects. A spatial 
    sample drift correction is, furthermore, performed after every forth row. 
    
  \item To get an appropriate S/N ratio, the data are preferentially 
    binned parallel to the interface. This is indicated with the
    green box in Fig.~\ref{fig:EELS_experiment}. The binning is done
    by a script introduced in the Gatan Microscopy Suite DigitalMicrograph 
    (v$2.32.888.0$, $32$-bit). The binned EEL spectra are processed 
    further in Digital Micrograph according to the following steps:
    Zero-loss centering using the reflected tail method, pre-edge
    background subtraction via an inverse power law function with 
    a fitting window from $407$~eV to $452$~eV and subsequent 
    deconvolution of plasmon contribution by the Fourier-ratio 
    method~\cite{Williams_2009}.
    
  \item A self developed MATLAB script is used for the further data evaluation.
    A Hartree-Slater cross section (exported from DigitalMicrograph for the 
    Ti $L$ edge) is subtracted from the Ti $L_{3,2}$ edge to compensate for
    the $L_3$-$L_2$ overlap. Fig.~\ref{fig:Ti-L-ELNES_with_fit} shows the 
    resulting Ti $L_{3\textrm{,}2}$ electron energy-loss near-edge structure 
    (ELNES). Five Lorentz functions are fitted to the five Ti
    $L_{3\textrm{,}2}$ ELNES peaks and the resulting envelope function
    is used to evaluate the physical parameters. An example of an evaluated
    parameter is the energy splittings d$E(b'-a)$.
    The parameters for the same $z$ position are finally evaluated to
    calculate the mean value and the standard deviation.
    
  \item Alternatively, all EEL spectra with the same $z$ distance to
    the interface can be binned and averaged to one EEL spectrum.
    This 'more classical' approach allows to easily calculate the
    difference spectra shown in Fig.~\ref{fig:energy_splitting} by the
    sacrifice of the error bar.
\end{enumerate}

In vertical direction the surface of the TEM-lamella was covered with 
Pd, allowing for dissociative chemisorption of the hydrogen gas molecules. In 
lateral direction the lamella offers two uncovered TiO$_2$ surfaces. Hydrogen entry 
from these sides is not expected.

\begin{figure}[!ht]
 \centering
\includegraphics[width=0.8\columnwidth]{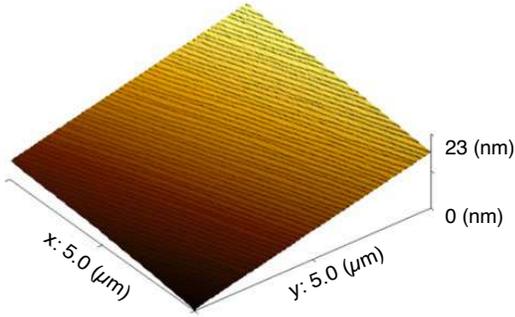}
 \caption{\label{fig:AFM_experiment}25 $\mu$m$^2$ 3D AFM height 
 profile of the r-TiO$_2$ crysal surface, after pretreatment. The surface 
 shows flat terraces of about 100 nm width with atomic steps inbetween.}
\end{figure}

It should be noted that the substrate was monitored by Atomic Force 
Microscope (AFM) after pretreatment and before Pd deposition (see 
Fig.~\ref{fig:AFM_experiment}), yielding clear flat terraces related to the 
substrate miscut. It shows about 50 atomic steps over 5 $\mu$m substrate length. 
This leads to a flat terrace width of 100 nm. The TEM lamella is of much smaller 
size and should not contain more than 2 terraces, meaning one atomic step.

\begin{figure}[!ht]
 \centering
\includegraphics[width=0.8\columnwidth]{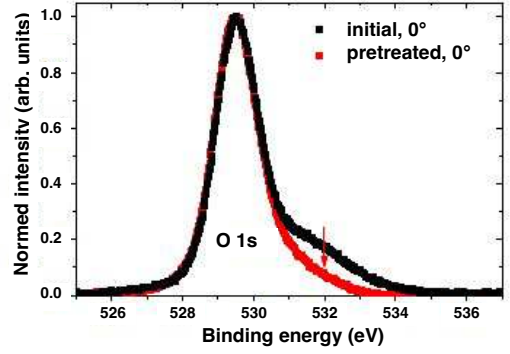}
 \caption{\label{fig:X-ray_experiment}X-ray photoelectron spectra of 
 the r-TiO$_2$ substrate surface at an incident angle of 0$^\circ$. 
 The oxygen 1s edge is shown before and after the pretreatment. The 
 oxygen 1s edge has a shoulder which decreased after the pretreatment.}
\end{figure}

X-ray photoemission (XPS) also was performed to verify the clean 
substrate after substrate pretreatment. The oxygen 1$s$ peak is shown in 
Fig.~\ref{fig:X-ray_experiment}. After pretreatment, it shows the expected 
spectrum for a clean surface. The initial surface contains contributions coming 
from C-O, C-H or H-O-H surface species.

\section{TMD model for defect concentrations}
\label{sec:model}
Defect concentrations (probabilities) can be calculated from the defect energies
including the configurational entropy of the defect distribution via the grand potentials 
of the bulk ($\Omega_1$) and at the interface ($\Omega_1$ and $\Omega_2$). The
electrostatic interaction is treated on a mean-field level by electrostatic potential of the 
defect charge densities. The details are described in the following. 

We consider a multistate model, where an oxygen site can be in one of
four states $X_{\sigma}$ (see Tab.~\ref{tab:pottstate}). These states can either 
be an oxygen atom or one of the defects in its respective charge state at
an oxygen site. The interstitial hydrogen is also attributed to 
an oxygen site. 

The grand potential $\Omega$ is a sum over all
microstates, each of which is described by a vector
$\vec{\sigma}$. Each component $\sigma_j$ of this vector selects the
state of the particular oxygen site at position $\vec{R_j}$ in the
r-TiO$_2$ lattice. The integer value $\sigma_j$ selects one of the possible
states $X_{\sigma}$ of an oxygen site. 
The grand potential of the defects is given by, 
\begin{eqnarray}
\Omega_1=-k_{\text{B}}T\ln\sum_{\vec{\sigma}}
\exp\left[-\beta\left(\mathcal{E}_{\vec{\sigma}}
-\hspace{-4mm}\sum_{k\in\{e,\text{O},\text{H}\}}\hspace{-2mm}
\mu_k\mathcal{N}_{k,\vec{\sigma}}\right)\right]
\nonumber\\
\end{eqnarray}
where $\beta=1/(k_{\text{B}}T)$. The system is coupled to particle reservoirs
with chemical potentials $\mu_e$ for electrons, $\mu_{\text{H}}$ for hydrogen,
and $\mu_{\mathrm{O}}$ for oxygen.

The total energies $\mathcal{E}_{\vec{\sigma}}$ and particle
numbers $\mathcal{N}_{\vec{\sigma}}$ of the defect distributions are
\begin{eqnarray}
\mathcal{E}_{\vec{\sigma}}
&=&\sum_j \left(E_{\sigma_j}+\eta_{e,\sigma_j}v_e(R_j)\right)
\nonumber\\
\mathcal{N}_{k,\vec{\sigma}}&=&\sum_j\eta_{k,\sigma_j}
\end{eqnarray}
where $\eta_{k,\sigma}$ with $k\in\{e,\text{O},\text{H}\}$ are particle numbers
for the specified defect as listed in Tab.~\ref{tab:pottstate}. The
particles considered are electrons, hydrogen atoms and oxygen atoms.

$v_e(\vec{r})=-e\Phi(\vec{r})$ is the potential acting on the
electrons due to the electric potential $\Phi(\vec{r})$.  It is
treated within the mean field theory: it plays the role of an external
potential, which is later connected self-consistently to the charge
density.

In addition, we consider the charge carriers like
electrons and holes which are not related to defects. We start with the grand potential of
non-interacting electrons with a density of states $D(\epsilon)$.
\begin{eqnarray}
\Omega_2&=&-k_{\text{B}}T
\int_{-\infty}^\infty d\epsilon\int d^3r\;
\frac{1}{V}D\left(\epsilon-v_e(\vec{r})\right)
\nonumber\\
&&\times
\ln\left[1+\exp\left(-\frac{|\epsilon-\mu_e|}{k_{\text{B}}T}\right)\right]
\label{eq:omega2}
\end{eqnarray}
The energy for the filled Fermi sea has been subtracted, which is
accounted for by taking the absolute value of $\epsilon-\mu_e$. The
expression is limited to Fermi-levels within the band-gap of the
oxide.

The functional form of the density of states
$D(\epsilon)=D_v(\epsilon)+D_c(\epsilon)$ is defined, like
conduction-band edge $\epsilon_c$ and the valence-band edge
$\epsilon_v$, in the absence of an electric potential. The latter is
accounted for by the choice of the argument. The density of states is
approximated by parabolic bands for valence and conduction bands with
multiplicities $N_v$ and $N_c$ and effective masses $m^*_v$ and
$m^*_c$.
\begin{eqnarray}
D_c(\epsilon)&=&2\pi V N_c\left(\frac{\sqrt{2m^*_c}}{2\pi\hbar}\right)^3 
  \sqrt{\epsilon-\epsilon_c} \quad\text{for $\epsilon>\epsilon_c$}
\nonumber\\
D_v(\epsilon)&=&2\pi V N_v\left(\frac{\sqrt{2m^*_v}}{2\pi\hbar}\right)^3 
  \sqrt{\epsilon_v-\epsilon}\quad\text{for $\epsilon<\epsilon_v$}
  \nonumber\\
\label{eq:dedv}
\end{eqnarray}
The volume $V$ in this expression is canceled in Eq.~\ref{eq:omega2}.
The effective masses have been set to $m_c^*=m_e$ for the
conduction-band minimum and $m_v^*=3.33~m_e$ for the valence-band
maximum~\cite{Perevalov_2011}. A one to two average of the effective masses 
calculated with GGA~\cite{perdew96_prl77_3865} for the $\Gamma \rightarrow Z$ 
and $\Gamma \rightarrow M/Z$ directions has been used. The multiplicities from 
our calculated band structure are $N_c=4$ for the conduction-band minimum 
and $N_v=2$ for the valence-band top, including spin multiplicity.

When evaluating the grand potential of free carriers, we retain only
the leading term in the Boltzmann factor
$\exp(-\beta|\epsilon-\mu_e|)$.

After adding the grand potentials of the defects and free charge
carriers, we obtain the grand potential $\Omega=\Omega_1+\Omega_2$ for
a given potential $v_e(\vec{r})$ as
\begin{eqnarray}
&&\Omega(T,\mu_e,\mu_{\text{H}},\mu_{\text{O}},[v_e])
=\int d^3r\;\biggl\lbrace
-\frac{k_{\text{B}}T}{\text{V}_{\text{O}}}
\nonumber\\
&\times&
\ln\left(\sum_{\sigma=1}^{N_{sp}}
\exp\Bigl[-\beta\Bigl(E_\sigma
+\eta_{e,\sigma}v_e(\vec{r})
-\hspace{-2mm}\sum_{k\in\{e,\text{O},\text{H}\}}\hspace{-2mm}
\eta_{k,\sigma}\mu_k\Bigr)\Bigr]\right)
\nonumber\\
&&-\left(\frac{m_e}{2\pi\hbar^2}\right)^{\frac{3}{2}}
(k_{\text{B}}T)^{\frac{5}{2}}
\Bigl[
N_c\left(\frac{m_c^*}{m_e}\right)^{\frac{3}{2}}
\e{-\beta(\epsilon_c+v_e(\vec{r})-\mu_e)}
\nonumber\\
&&\hspace{2cm}
+N_v\left(\frac{m_v^*}{m_e}\right)^{\frac{3}{2}}
\e{-\beta(\mu_e-\epsilon_v-v_e(\vec{r}))}
\Bigr]\biggr\rbrace
\label{eq:tmdgrandpot}
\end{eqnarray}
Here
$\text{V}_{\text{O}}\approx1.5\times10^{-23}$~cm$^3$ is the volume per oxygen 
site in
r-TiO$_2$.

Within our model, the grand potential contains the complete
thermodynamic information of the problem.  The local particle
concentrations (particles per volume) $n_k(\vec{r})$ are obtained, up
to a sign change, as the functional derivative of the grand potential
with respect to a local variation of the respective chemical
potential, namely
\begin{eqnarray}
\label{eq:parcon}
n_k(\vec{r})=-\frac{\delta\Omega}{\delta\mu_k(\vec{r})}
=\frac{1}{\text{V}_{\text{O}}}\sum_{\sigma}P_\sigma(\vec{r})\eta_{k,\sigma}
\end{eqnarray}
for $k\in\{\text{H},\text{O}\}$. For the electrons we need to add the 
contribution of the free charge carriers.
\begin{eqnarray}
n_e(\vec{r})&=&-\frac{\delta\Omega}{\delta\mu_e(\vec{r})}
=\frac{1}{\text{V}_{\text{O}}}\sum_{\sigma}P_\sigma(\vec{r})\eta_{e,\sigma}
+\left(\frac{m_ek_{\text{B}}T}{2\pi\hbar^2}\right)^{\frac{3}{2}}
\nonumber\\
&\times&
\biggl[
N_c\left(\frac{m_c^*}{m_e}\right)^{\frac{3}{2}}
\e{-\beta(\epsilon_c+v_e(\vec{r})-\mu_e)}
\nonumber\\
&-&
N_v\left(\frac{m_v^*}{m_e}\right)^{\frac{3}{2}}
\e{-\beta(\mu_e-\epsilon_v-v_e(\vec{r}))}
\biggr]
\;.
\label{eq:electrondensity}
\end{eqnarray}

The probability $P_\sigma(\vec{r})$ that an oxygen site at position
$\vec{r}$ is in state $\sigma$, is, according to
Eqs.~\ref{eq:tmdgrandpot} and \ref{eq:parcon},
\begin{eqnarray}
P_\sigma(\vec{r})=
\frac{
\exp\left(
-\frac{1}{k_{\text{B}}T}(E_\sigma+\eta_{e,\sigma}v_e(\vec{r})-
\sum_{k}\eta_k\mu_k)
\right)
}{\sum_{\sigma=1}^{4}
  \exp\left(
-\frac{1}{k_{\text{B}}T}(E_\sigma+\eta_{e,\sigma}v_e(\vec{r})-
\sum_{k}\eta_k\mu_k)
\right)}\nonumber\\
\label{eq:probability}
\end{eqnarray}

The charge density is $\rho(\vec{r})=-e n_e(\vec{r})$. The electric
potential $\Phi(\vec{r})$ is connected with the potential acting on
the electrons by $v_e(\vec{r})=-e\Phi(\vec{r})$. The Poisson equation
closes the system of self-consistent equations
\begin{eqnarray}
\vec{\nabla}^2\Phi(\vec{r})=-\frac{1}{\epsilon_0\epsilon_r}\rho(\vec{r})
\label{eq:poisson}
\end{eqnarray}
The relative dielectric constant of $\epsilon_r=86$ has been taken
from Parker~\cite{Parker_1961}.

For the description of the interface, the  
electric potential is allowed to vary only perpendicular to the
interface. The distance from the interface is denoted by $z$, with
the positive direction pointing towards the oxide.
With $\vec{r}_I$, we denote a position at the interface and
$\vec{r}_B$ is a position deep in the bulk oxide.  The potential at
the interface is thus $v_e(\vec{r}_I)=v_e(z=0)$ and that in the
bulk is $v_e(\vec{r}_B)=v_e(z=+\infty)$.

The alignment of the band structures at the interface is determined by
the Schottky-barrier height $\Phi_{\text{Bn}}=\epsilon_c+v(\vec{r}_I)-\mu_e$,
defined as the difference of the conduction-band edge
$\epsilon_c+v_e(\vec{r}_I)$ at the interface from the Fermi level
$\mu_e$ of the metal. We have chosen its value
$\Phi_{\text{Bn}}=0.79$~eV~\cite{DuBridge_1932, Imanishi_2007} from the
Schottky model as the metal work function minus the electron affinity
of the oxide. Note that the interface between metal and oxides is 
affected by the local chemistry and the Schottky barrier. The Schottky barrier 
height alters for different terminations and has the dominant influence on the 
stability of impurities and dopants at the interface rather than the local 
chemistry.

In the oxide far from the interface, the position of the electron
chemical potential $\mu_e$ relative to the band edges is determined by
the charge-neutrality condition. Thus, the electric potential
$v_e(\vec{r}_B)$ far from the interface relative to the electron
chemical potential is a bulk property, which, however, depends on the
chemical potentials of oxygen and hydrogen.

Near the interface, charges build up in the oxide, which produce an
electric-potential step that aligns the charge-neutrality level of the
bulk oxide with the Fermi level of the metal. The global charge
neutrality is satisfied by a neutralizing charge buildup in the metal
right at the interface. The charge density of the interface system can
be expressed as
\begin{eqnarray}
\rho(z)=
-e\Bigl[\theta(z)n_e(z)-\delta(z-0^-)\int_{0^+}^\infty dz\; n_e(\vec{r})
\Bigr]\nonumber\\
\end{eqnarray}
with the the step function $\theta(z)$ and the delta-function
$\delta(z)$.  This charge density defines the electron potential via
the Poisson equation.
\begin{eqnarray}
v_e(z)&=&\mu_e+\frac{e}{\epsilon_0\epsilon_r}\delta(z-0^-)\Big[z\Big(\int_{0^+}^\infty dz'n_e(z')\Big)
\nonumber\\
&-&
\int_{0}^z dz \int_{0}^{z'} dz'' n_e(z'') \Big]
\label{eq:charge-distribution}
\end{eqnarray}
Equations~\ref{eq:electrondensity},~\ref{eq:probability},
and~\ref{eq:charge-distribution} form a coupled set of equations, that
needs to be solved selfconsistently.

In the self consistent calculation of the potential profile has not
been required, because the potential step $\Delta{v}_e$ between the
interface and the bulk could be obtained from the Schottky barrier.

This concludes the description of the TMD model. The parameters used
in the model are summarized in tables~\ref{tab:pottstate} and
\ref{tab:modelparms}.

\begin{table}[!hbt]
\caption{\label{tab:modelparms}Parameters used for the model of defect
  concentrations.}
\begin{center}
\begin{tabular}{llrr}
\hline
\hline
effective electron mass &$m_c^*$ & $m_e$ & Ref.~\cite{Perevalov_2011}\\
effective hole mass & $m_v^*$ & 3.33~$m_e$ & Ref.~\cite{Perevalov_2011}\\
conduction band multiplicity & $N_c$ & 4 & \\
valence band multiplicity & $N_v$ & 2 & \\
volume per oxygen site (cm$^3$) &   $V_{\text{O}}$ & $1.5\times10^{-23}$ & \\
relative dielectric constant    & $\epsilon_r$ &86 & Ref.~\cite{Parker_1961} \\
\gga{PBE} band-gap (eV) & $\epsilon_c-\epsilon_v$ & 
\gga{1.90} & \\
Schottky barrier (eV)    & $\Phi_{\text{Bn}}$ & 0.79 & Ref.~\cite{Bongers_2018}\\
\hline
\multicolumn{2}{l}{
$k_{\text{B}}T\ln\left[N_c\text{V}_{\text{O}}
\left(m_c^*k_{\text{B}}T/(2\pi\hbar^2)\right)^{\frac{3}{2}}\right]$
} & -1.62~eV\\
\hline
\hline
\end{tabular}
\end{center}
\end{table}

\section{Partial pressures and chemical potentials}
\label{app:partpress}
In order to link the TMD model to measurable quantities, we need to
convert the partial pressures of the gases O$_2$ and H$_2$ in the
environment into chemical potentials for the atoms.  The resulting
chemical potentials for oxygen and hydrogen obtained for the
individual processing steps are summarized in Tab.~\ref{tab:chempots}.

For dimeric molecules, the chemical potentials $\mu_{Y}$ for the
atomic species, $Y\in\{\textrm{O},\textrm{H}\}$, are obtained from the
Gibbs free energy $G_{Y_2}(T,p,N)$ of the corresponding gas $Y_2$ at
the specified partial pressure $p$ and temperature $T$.  The chemical
potentials $\mu_{Y_2}=G_{Y_2}/N_{Y_2}$ for the molecules are obtained
as the Gibbs free energy per molecule. The chemical potential of an
atom $\mu_Y=\frac{1}{2}\mu_{Y_2}$ is one-half of that of the
dimeric molecule. Thus, we obtain
\begin{eqnarray}
\mu_{Y}=\frac{1}{2N}G_{Y_2}(T,p,N)\;.
\label{eq:chemicalpotentialfrompartialpressure}
\end{eqnarray}

The molecular Gibbs free energy (dropping the subscript $Y_2$)
\begin{eqnarray}
G(T,p,N)&=&
N E_{\text{tot}}+G_{trans}(T,p,N)+F_{vib}(T,N)
\nonumber\\
&&+F_{rot}(T,N)+F_{e}(T,N)
\end{eqnarray}
is divided into the total
energy $E_{\text{tot}}$ of the molecules without their zero-point energy,
its contribution $G_{\text{trans}}$ from molecular translations, the
free energy $F_{\text{vib}}$ of internal vibration including the
zero-point energy, the free energy $F_{\text{rot}}$ of the molecular
rotations, as well as the contribution from spin multiplicity,

The translational contribution to the Gibbs potential
\begin{eqnarray}
G_{trans}(T,p,N)=Nk_{\text{B}}T\ln\left(\frac{\lambda_T^3p}{k_{\text{B}}T}\right)
\end{eqnarray}
depends on the thermal de Broglie wave length
$\lambda_T=\sqrt{\frac{2\pi\hbar^2}{Mk_{\text{B}}T}}$, where $M$ is
the mass of the molecule.  As we consider a dilute gas of molecules,
the interaction between molecules is ignored, so that we use the Gibbs
free energy of the ideal gas.

The vibrational free energy is expressed by the vibrational frequency
$\omega$ as
\begin{eqnarray}
F_{vib}(T,N)=\frac{N}{2}\hbar\omega
+Nk_BT\ln\Bigl[1-\exp\Bigl(-\frac{\hbar\omega}{k_BT}\Bigr)\Bigr]\;.
\end{eqnarray}
It contains the zero point energy as well as the quantum mechanical
entropy term.

The rotational part is obtained 
\begin{eqnarray}
F_{\text{rot}}(T,N)&=&-Nk_BT\ln
\biggl[
\sum_{\ell=0}^{\infty}(2\ell+1){\rm e}^{-\frac{\hbar^2\ell(\ell+2)}{\mu d^2 k_BT}}
\biggr]
\end{eqnarray}
The effective mass $\mu$ of the dimer-molecule is one-half of the
atomic masses, respectively one fourth of the molecular mass $M$,
i.e. $\mu=\frac{1}{4}M$. The bond distance is denoted as $d$.

For the oxygen molecule also the spin multiplicity $g_e=3$ contributes to
the free energy in the form
\begin{eqnarray}
F_{e}(T,N)=-Nk_BT\ln(g_e)
\end{eqnarray}

The molecular parameters entering
Eq.~\ref{eq:chemicalpotentialfrompartialpressure} are summarized in
Tab.~\ref{tab:molpar}.
\begin{table}[!hbt]
\caption{\label{tab:molpar}Molecular parameters~\cite{Shimanouchi72}
 used to convert temperatures and partial pressures via
  Eq.~\ref{eq:chemicalpotentialfrompartialpressure} into chemical
  potentials.
}
\begin{center}
\begin{tabular}{|l|c|c|c|}
\hline
       & $g_e$ & $d$     & $\hbar\omega/(2\pi\hbar c)$         \\
\hline
O$_2$  &  3 & 1.208~\AA & 1580.19~cm$^{-1}$ \\
H$_2$  &  1 & 0.741~\AA & 4163.17~cm$^{-1}$ \\
 \hline
\end{tabular}
\end{center}
\end{table}

%

\end{document}